\documentclass[11pt,a4paper]{article}
\pdfoutput=1

\usepackage{amsmath}
\usepackage[T1]{fontenc}

\usepackage{jheppub}
\usepackage{psfrag}
\usepackage{slashed}
\usepackage{cancel}
\usepackage{lscape}
\usepackage{float}
\usepackage{caption}
\usepackage{array}
\usepackage{graphicx}
\usepackage{subcaption}
\usepackage{multirow}
\usepackage{tabularx}
\usepackage{makecell}
\usepackage[export]{adjustbox}

\usepackage[utf8]{inputenc}

\def\ttgg{\bar{t}tgg}
\def\tb{{\bar{t}}}
\def\eps{\epsilon}
\def\nn{\nonumber \\ }
\def\mren{\mathrm{mren}}
\def\ren{\mathrm{ren}}
\def\mct{\mathrm{mct}}
\def\cA{\mathcal{A}}
\def\cO{\mathcal{O}}
\def\cJ{\mathcal{J}}
\def\cI{\mathcal{I}}
\def\cT{\mathcal{T}}
\def\zz{\boldsymbol{Z}}
\def\mi{\mathrm{MI}}
\def\fin{\mathrm{fin}}
\def\as{\alpha_s}
\def\dk#1{\frac{d^d k_{#1}}{i\pi^{d/2}e^{-\eps \gamma_E}}}
\def\ceps{C_\eps}

\def\la{\langle}
\def\ra{\rangle}
\def\spA#1#2{\la#1#2\ra}
\def\spB#1#2{[#1#2]}
\def\spAB#1#2#3{\la#1|#2|#3]}
\def\spAA#1#2#3{\la#1|#2|#3\ra}
\def\spBB#1#2#3{[#1|#2|#3]}
\def\spAXB#1#2#3#4#5{\la#1|#2|#3|#4|#5]}
\def\fl#1{#1^\flat}

\def\usepix#1#2#3#4#5#6{\parbox{#1}{\includegraphics[width=#1,trim= #3 #4 #5 #6,clip=true]{#2}}}

\newcolumntype{C}[1]{>{\hsize=#1\hsize\centering\arraybackslash}X}%

\newcolumntype{Z}{r<{\hspace{3mm}}}
\newcommand{\ba}{\[\begin{aligned}}
\newcommand{\ea}{\end{aligned}\]}

\newcommand{\curveone}{(a)}
\newcommand{\curvetwo}{(b)}
\newcommand{\curvethree}{(c)}

\title{
Two-loop leading colour QCD helicity amplitudes for top quark pair production in the gluon fusion channel
}

\author[a]{Simon Badger,}
\author[a]{Ekta Chaubey,}
\author[b]{Heribertus Bayu Hartanto,}
\author[c]{Robin Marzucca}

\affiliation[a]{
Dipartimento di Fisica and Arnold-Regge Center, Universit\`{a} di Torino,
and INFN, Sezione di Torino, Via P. Giuria 1, I-10125 Torino, Italy
}
\affiliation[b]{
Cavendish Laboratory, University of Cambridge, Cambridge CB3 0HE, United Kingdom
}
\affiliation[c]{
Institute for Particle Physics Phenomenology, Department of Physics, Durham University, Durham DH1
3LE, United Kingdom%
}

\emailAdd{
simondavid.badger@unito.it,
hbhartanto@hep.phy.cam.ac.uk,
ekta.xxx@unito.it,
robin.marzucca@durham.ac.uk
}

\abstract{
We present a complete set of analytic helicity amplitudes for top quark pair production via gluon
fusion at two-loops in QCD. For the first time, we include corrections due to massive fermion loops
which give rise to integrals over elliptic curves. We present the results of the missing master integrals
needed to compute the amplitude and obtain an analytic form for the finite remainders in terms of
iterated integrals using rationalised kinematics and finite field sampling. We also study the
numerical evaluation of the iterated integrals.
}

\preprint{CAVENDISH-HEP-21/04}

\begin{document}
\maketitle
\flushbottom
\section{Introduction \label{sec:intro}}

Precise understanding of the top quark in the Standard Model is one of the highest priorities in
modern collider experiments. The large mass of the top quark makes it a dominant ingredient in
understanding the fundamental forces and in particular of the electroweak symmetry breaking
mechanism. A precise determination of the top quark mass is imperative due to the sensitivity of the Higgs
potential on its value.

The QCD corrections to top-quark pair production have a long history going back to the first
total cross section computation at NLO for on-shell top-quarks computed by Nason, Dawson and Ellis~\cite{Nason:1987xz,Nason:1989zy}.
Thanks to a monumental effort on the part of the theoretical community, fully differential NNLO predictions for
top anti-top pair production at hadron colliders are now available for comparisons with the latest
experimental data~\cite{Baernreuther:2012ws,Czakon:2012zr,Czakon:2012pz,Czakon:2013goa,Czakon:2015owf}. These intensive computations require two-loop $2\to 2$, one-loop $2\to 3$ and
tree-level $2\to4$ partonic amplitudes to be combined using a consistent UV and IR subtraction
scheme to obtain a finite result. The first complete computation used the `STRIPPER' sector
decomposition method to subtract IR divergences and has enabled the most complete theoretical
descriptions of top-quark dynamics~\cite{Czakon:2010td}. Very recently a second independent computation has been obtained
using the $q_T$ subtraction method~\cite{Bonciani:2015sha,Catani:2019iny,Catani:2019hip,Catani:2020tko}.

Up to now the only complete two-loop amplitudes have been available
numerically~\cite{Czakon:2008zk,Baernreuther:2013caa,Chen:2017jvi}. The main reason for
this is that a more complicated class of special functions begin to appear in amplitudes which
contain internal masses. These functions have been identified to involve integrals over elliptic curves ~\cite{Adams_2017,Adams:2018bsn,Adams:2018kez,adams2018feynman,Abreu_2020,Adams_2018,BOGNER2017528,Broedel:2019kmn,Abreu:2019fgk}, and lie on the boundary of our current mathematical understanding. Over the last few years,
motivated by the increasing demand of precise theoretical descriptions including mass dependence,
there has been substantial progress in developing a complete framework for use in phenomenological
applications. There are a large number of amplitude level corrections to the process which do not
depend on elliptic sectors and there has been substantial effort to find a complete analytic
form for the squared
amplitudes~\cite{Bonciani:2008az,Bonciani:2009nb,Bonciani:2010mn,Bonciani:2013ywa,vonManteuffel:2013uoa,DiVita:2018nnh,Becchetti:2019tjy,DiVita:2019lpl}.

In this paper we present a set of helicity amplitudes for top-quark pair production in the leading
colour approximation. For the first time we also include contributions from heavy fermion loops
which lead to the appearance of iterated integrals involving elliptic curves. The master integrals for these heavy loop
corrections have been recently completed~\cite{Adams:2018kez,Adams:2018bsn}. As well as obtaining compact analytic helicity
amplitudes by sampling Feynman diagrams with finite field
arithmetic~\cite{Wang:1981:PAU:800206.806398,Wang:1982:PRR:1089292.1089293,Trager:2006:1145768,vonManteuffel:2014ixa,Peraro:2016wsq,Peraro:2019svx}, we also study the numerical
evaluation of the final amplitudes. The helicity amplitudes presented contain complete information
about top quark decays in the narrow width approximation. This helicity amplitude technique has been
used successfully at one-loop~\cite{Ellis:2008ir,Melnikov:2009dn,Badger:2011yu,Badger:2017gta}. Spin
correlations can also be treated efficiently using the spin density matrix
approach~\cite{Bernreuther:2001rq,Bernreuther:2004jv,Chen:2017jvi}

As is always the case, computations at this perturbative order contain a large number of steps, each
usually with some technical bottlenecks to overcome. The analytic computations of the planar master
integrals, including those with an internal massive loop, are available in the literature for almost
every topology. Yet, for the construction of the amplitude it was necessary to include one final two-loop integral topology which
contained two elliptic master integrals. The (canonical form) differential
equation~\cite{Bern:1993kr,Kotikov:1990kg,Remiddi:1997ny,Gehrmann:1999as,Henn:2013pwa,Adams_2018} and solution in terms of iterated integrals was obtained and combined with
the integration-by-parts reduced helicity amplitudes.

Throughout the computation we made use of finite field arithmetic to find an efficient solution to
the system of integration-by-parts identities. These techniques have shown to be particularly
efficient for massless amplitudes with many external scales. In this paper we show how they can
apply equally well to amplitudes with massive internal particles. We construct a rational
parametrisation of the on-shell kinematics via momentum twistors and obtain helicity amplitudes via
projecting onto a basis of independent spinor structures, accounting for the freedom in the top quark
spin states. This method defines a set of on-shell, gauge invariant sub-amplitudes that can be
computed using on-shell top-quark kinematics. We then check these results against previous
computations for the interference with tree-level diagrams in the conventional dimensional regularisation (CDR) scheme.

\section{Leading colour $\ttgg$ amplitudes}

We consider a scattering process involving a pair of top quarks and two gluons
\begin{equation}
0 \rightarrow \bar{t}(p_1) + t(p_2) + g(p_3) + g(p_4), \nonumber
\end{equation}
where $p_1^2 = p_2^2 = m_t^2$ and $p_3^2 = p_4^3 = 0$. The kinematic invariants for this process 
are the top-quark mass $m_t$, and the two Mandelstam variables
\begin{equation}
  s=(p_1+p_2)^2, \qquad t=(p_2+p_3)^2.
\end{equation}

In this work we consider the leading colour contributions of the $\ttgg$ amplitude up to two-loop level, where at two loops,
only planar configurations arise. The colour decomposition of the leading colour $L$-loop $\ttgg$ amplitude is given by
\begin{align}
\mathcal{A}^{(L)}(1_{\tb}, 2_{t}, 3_{g}, 4_{g}) &= n^L g_s^2 
 \; \bigg[
 \left( T^{a_{3}} T^{a_{4}} \right)_{i_2}^{\;\;\bar i_1}
 A^{(L)}(1_\tb,2_t, 3_g ,4_g) + (3 \leftrightarrow 4)
 \bigg],
\label{eq:colourdecomposition}
\end{align}
where $n= m_\eps \alpha_s/(4\pi),\ \alpha_s = g_s^2/(4\pi)$, $m_\eps=i \left(4\pi/m_t^2\right)^{\eps} e^{-\eps\gamma_E}$, $g_s$ is the strong coupling constant
and $(T^a)_i^{\;\;\bar j}$ are the fundamental generators of $SU(N_c)$.

\begin{figure}[t]
  \begin{center}
    \includegraphics[width=0.8\textwidth]{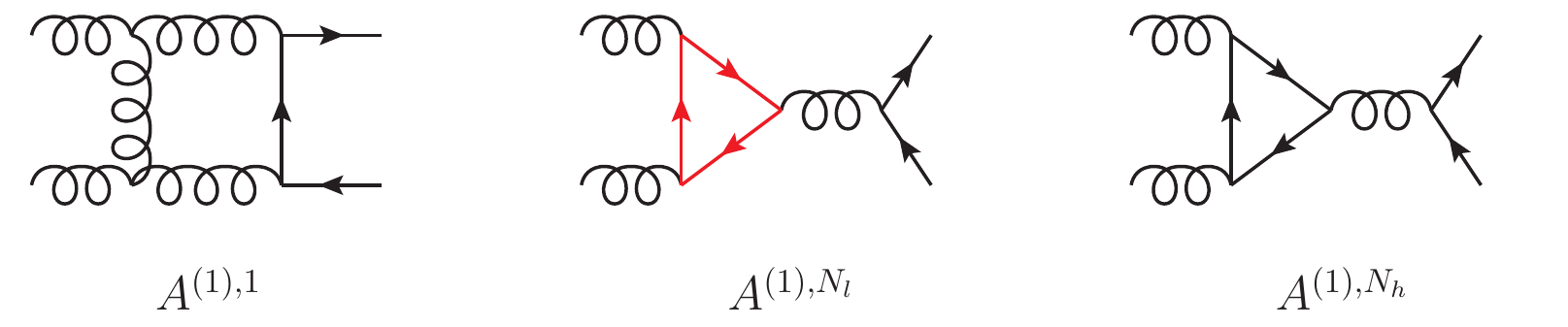} 
  \end{center}
  \caption{Sample Feynman diagrams corresponding to various internal flavour contributions at one loop as specified in Eq.~\eqref{eq:flamplitude1loop}. 
   Red lines, black spiral lines and black lines represent massless quarks, gluons and top quarks, respectively.}
  \label{fig:diagrams1L}
\end{figure}

\begin{figure}[t]
  \begin{center}
    \includegraphics[width=0.85\textwidth]{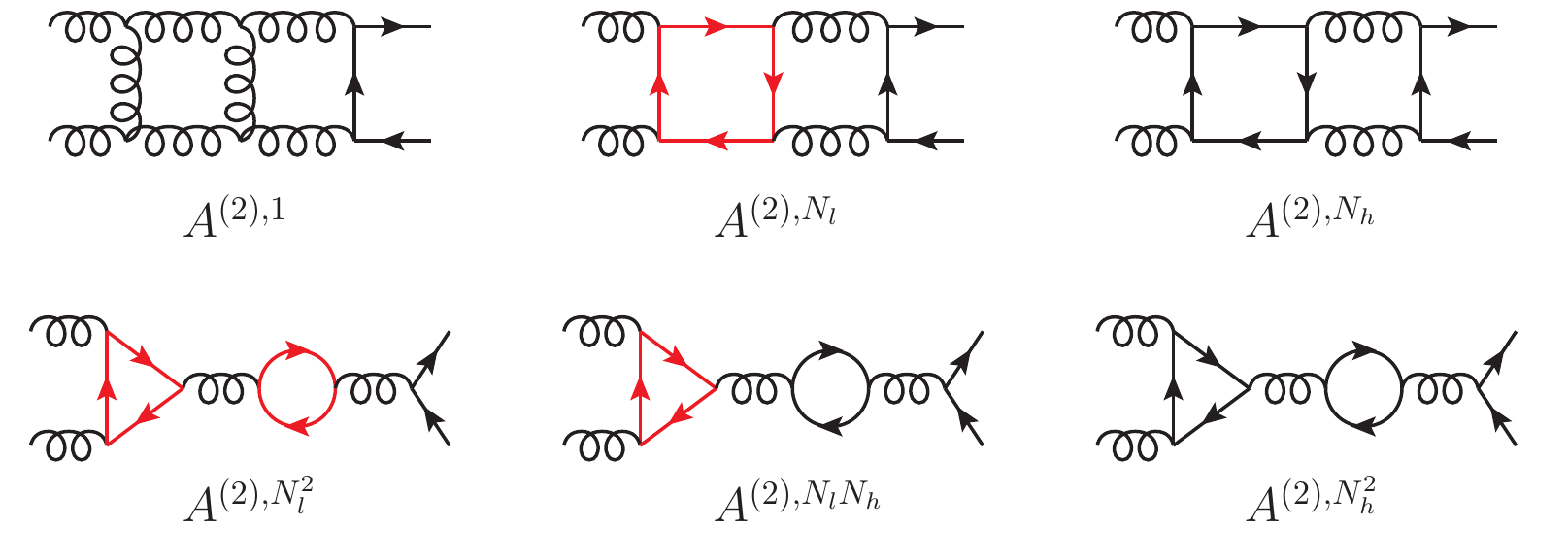} 
  \end{center}
  \caption{Sample Feynman diagrams corresponding to various internal flavour contributions at two loops as specified in Eq.~\eqref{eq:flamplitude2loop}. 
   Red lines, black spiral lines and black lines represent massless quarks, gluons and top quarks, respectively.}
  \label{fig:diagrams2L}
\end{figure}

\begin{figure}[t]
  \begin{center}
    \includegraphics[width=0.85\textwidth]{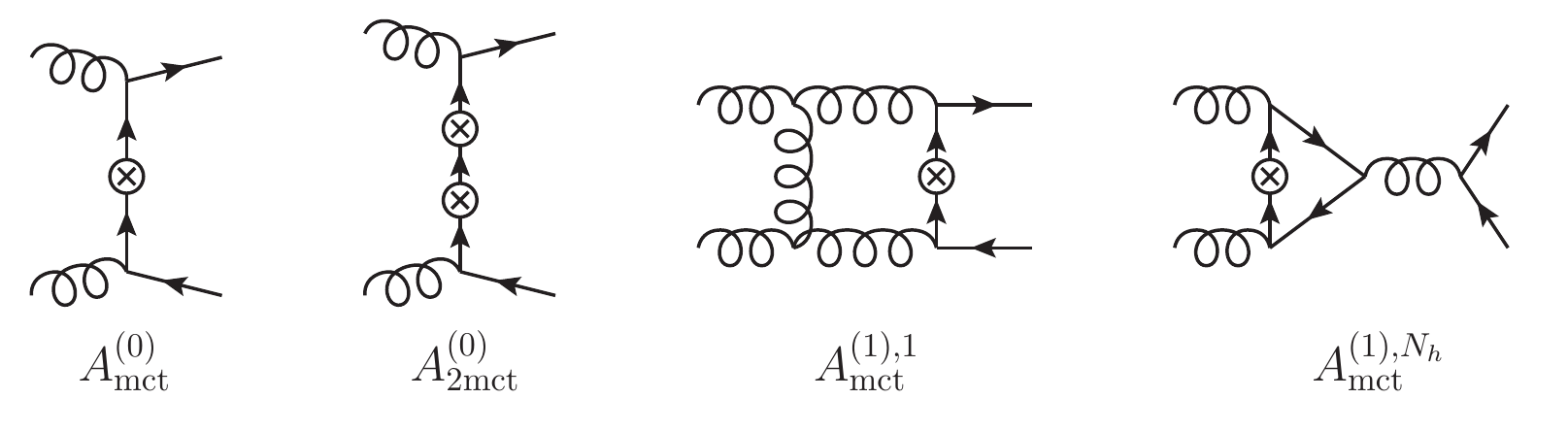} 
  \end{center}
  \caption{Sample Feynman diagrams with mass counterterm insertions at tree level and one loop that contribute 
  in Eqs.~\eqref{eq:mren_1loop}~and~\eqref{eq:mren_2loop}. 
  Black spiral lines and black lines represent gluons and top quarks, respectively.
  Circled crosses indicate the mass counterterm insertions.}
  \label{fig:diagramsMCT}
\end{figure}

The partial amplitudes can further be decomposed according to their internal flavour structure,
\begin{align}
A^{(1)}(1_\tb,2_t, 3_g ,4_g) = \; & N_c A^{(1),1} + N_l A^{(1),N_l} + N_h A^{(1),N_h},
\label{eq:flamplitude1loop} \\
A^{(2)}(1_\tb,2_t, 3_g ,4_g) = \; & N_c^2 A^{(2),1} + N_c N_l A^{(2),N_l} + N_c N_h A^{(2),N_h} \nn
& + N_l^2 A^{(2),N_l^2} + N_l N_h A^{(2),N_l N_h} + N_h^2 A^{(2),N_h^2}.
\label{eq:flamplitude2loop} 
\end{align}
where $N_l$ and $N_h$ are the number of  closed light and heavy quark loops, respectively. 
Sample diagrams for various fermion loop contributions at one and two loops are shown in Figs.~\ref{fig:diagrams1L}~and~\ref{fig:diagrams2L}.

We first define the gauge invariant $L$-loop mass-renormalised amplitude $A^{(L)}_\mren$, where the top-quark mass renormalisation is included
\begin{align}
A^{(1)}_\mren & = A^{(1)} - \delta Z_m^{(1)} A^{(0)}_\mct,
\label{eq:mren_1loop} \\
A^{(2)}_\mren & = A^{(2)} - \delta Z_m^{(2)} A^{(0)}_\mct - \delta Z_m^{(1)} A^{(1)}_\mct  +  \big(\delta Z_m^{(1)}\big)^2 A^{(0)}_{2\mct}.
\label{eq:mren_2loop}
\end{align}
$\delta Z_m^{(L)}$ is the $L$-loop mass renormalisation counterterm, $A^{(0)}_\mct(A^{(1)}_\mct$) represents an amplitude including one mass counterterm insertion to  
top-quark propagators in the tree level (one-loop) diagrams, while $A^{(0)}_{2\mct}$ corresponds to the tree level amplitude with two mass counterterm insertions 
to top-quark propagators. The mass counterterm insertion procedure leads to the presence of double and triple top-quark propagators in the tree level amplitude and 
double top-quark propagators in the one-loop amplitude. 
Sample diagrams with mass counterterm insertions at tree level and one loop are displayed in Fig.~\ref{fig:diagramsMCT}.

The fully renormalised $\ttgg$ amplitude can be obtained by including renormalisations of the top-quark and gluon wavefunctions as well as the strong coupling $\alpha_s$
\begin{align}
A^{(1)}_\ren = \; & A^{(1)}_\mren + \big( \delta Z^{(1)}_{t} + \delta Z^{(1)}_{g} + \delta Z^{(1)}_{\alpha_s} \big) A^{(0)}, 
\label{eq:ampren_1loop} \\
A^{(2)}_\ren = \; & A^{(2)}_\mren + \big( \delta Z^{(1)}_{t} + \delta Z^{(1)}_{g} + 2\delta Z^{(1)}_{\alpha_s} \big) A^{(1)}_\mren \nn
             &  + \big( \delta Z^{(2)}_{t} + \delta Z^{(2)}_{g}  + \delta Z^{(2)}_{\alpha_s} 
                + \delta Z^{(1)}_{t}\delta Z^{(1)}_{g} + 2\delta Z^{(1)}_{t}\delta Z^{(1)}_{\alpha_s} 
                + 2\delta Z^{(1)}_{g} \delta Z^{(1)}_{\alpha_s}   \big) A^{(0)}.
\label{eq:ampren_2loop} 
\end{align}
$\delta Z^{(L)}_{t}$, $\delta Z^{(L)}_{g}$ and $\delta Z^{(L)}_{\alpha_s}$ are the $L$-loop top-quark wavefunction, gluon wavefunction 
and strong coupling renormalisation constants, respectively. All the renormalisation constants relevant for our calculation are given 
in Appendix~\ref{app:renormalisation}.

The renormalisation procedure removes the ultraviolet (UV) singularities present in loop amplitudes.
The remaining divergences that are of infrared (IR) nature can be predicted from universal IR behaviour 
of QCD amplitudes~\cite{Catani:1998bh,Becher:2009cu,Becher:2009qa,Gardi:2009qi,Gardi:2009zv},
which, in the presence of top quarks, needs to be extended to massive case~\cite{Catani:2000ef,Ferroglia:2009ep,Ferroglia:2009ii}.
The UV renormalised amplitude at one and two loops can be divided into singular and finite terms
\begin{align}
|\cA_{n}^{(1)}\rangle & = \zz^{(1)} |\cA_{n}^{(0)}\rangle +  |\cA_{n}^{(1),\fin}\rangle,
\label{eq:amprenorm1loop} \\
|\cA_{n}^{(2)}\rangle & = \zz^{(2)} |\cA_{n}^{(0)}\rangle + \zz^{(1)} |\cA_{n}^{(1),\fin}\rangle + |\cA_{n}^{(2),\fin}\rangle.
\label{eq:amprenorm2loop} 
\end{align}
Here $|\cA_{n}^{(L)}\rangle$ is a vector in colour space and admits the following expansion in the strong coupling
\begin{equation}
|\cA_{n}\rangle = g_s^{n-2} \; \bigg\lbrace |\cA_{n}^{(0)}\rangle + \frac{\as}{4\pi} |\cA_{n}^{(1)}\rangle 
+ \left(\frac{\as}{4\pi}\right)^2 |\cA_{n}^{(1)}\rangle + \cO(\as^3) \bigg\rbrace.
\end{equation}
We remind that for the leading colour $\ttgg$ amplitudes the colour decomposition is given in Eq.~\eqref{eq:colourdecomposition}.
The IR divergences of the amplitude are encoded in the $\zz$ factor (which is a matrix in colour space) 
and up to two-loop order. For the process we are considering, the $\zz$ factor is given by~\cite{Ferroglia:2009ii}
\begin{align}
\zz = \; & 1 + \frac{\as}{4\pi} \bigg( \frac{\Gamma_0^\prime}{4\eps^2} + \frac{\boldsymbol\Gamma_0}{2\eps}  \bigg) \nonumber \\
& + \left(\frac{\as}{4\pi}\right)^2 \bigg\lbrace 
       \frac{(\Gamma_0^\prime)^2}{32\eps^4} 
     + \frac{\Gamma_0^\prime}{8\eps^3}\bigg( \boldsymbol\Gamma_0 - \frac{3}{2} \beta_0 \bigg) 
     + \frac{\boldsymbol\Gamma_0}{8\eps^2} \big( \boldsymbol\Gamma_0 - 2\beta_0 \big)
     + \frac{\Gamma_1^\prime}{16\eps^2}
     + \frac{\boldsymbol\Gamma_1}{4\eps} 
      \\
& - \frac{2 T_F}{3} \sum_{i=1}^{n_h} \bigg[ 
         \Gamma_0^\prime \bigg( \frac{1}{2\eps^2}\ln\frac{\mu^2}{m_i^2} + \frac{1}{4\eps}\bigg[\ln^2\frac{\mu^2}{m_i^2} +\frac{\pi^2}{6}\bigg] \bigg)
       + \frac{\boldsymbol\Gamma_0}{\eps}\ln\frac{\mu^2}{m_i^2} \bigg] \bigg\rbrace + \cO(\as^3). \nonumber
\end{align}
In this work we will consider the top quark as the only massive fermion, therefore $n_h=1$ and $m_1 = m_t$.
The coefficients $\boldsymbol\Gamma_n$ and $\Gamma^\prime_n$ are defined through the following expansion in $\alpha_s$,
\begin{equation}
\boldsymbol\Gamma = \sum_{n \geq 0} \boldsymbol\Gamma_n \bigg(\frac{\as}{4\pi}\bigg)^{n+1} 
\qquad \mathrm{and} \qquad 
\Gamma^\prime = -2 C_i \gamma_\mathrm{cusp}(\as).
\end{equation}
For the $\ttgg$ amplitude, $C_i = C_A = N_c$. 
The anomalous dimension matrix $\boldsymbol\Gamma$ for the $\ttgg$ amplitude can be obtained from Eq.~(55) of~\cite{Ferroglia:2009ii}.
We also refer the reader to Appendix~A of~\cite{Becher:2009qa} for the definition of anomalous dimension factors required 
for the computation of the IR poles of the $\ttgg$ amplitude.

\section{Helicity amplitudes for massive fermions}
\label{sec:massivehelicityamps}

There are a variety of different methods on the market for dealing with the spin states of a massive
fermion. The common aim is to find a compact notation at the amplitude level and retain all
information required to account for decays in the narrow width approximation. One major issue is
that the helicity for a massive particle is not conserved and so we must introduce an arbitrary
additional direction in order to define it. This additional extra direction can increase the
algebraic complexity of an analytic computation.

In this section we review one way in which massive spinors can be incorporated into the
spinor-helicity methods~\cite{Kleiss:1985yh} and then describe how to define a set of gauge invariant, on-shell sub-amplitudes
that can be used to describe the full set of spin correlated narrow width decays. These on-shell
sub-amplitudes can be computed using a rational parametrisation of the external kinematics. In our
case, this rational paramtrisation was generated using the momentum twistor formalism~\cite{Hodges:2009hk}.
The spinor-helicity method applied to massive fermions has been well studied and we refer the reader
to Refs.~\cite{Rodrigo:2005eu,Schwinn:2007ee,Arkani-Hamed:2017jhn} for further details and other approaches.

The first step is to introduce an arbitrary direction $n$ which can be used to define a massless
projection of the massive fermion momentum,
\begin{equation}
  p^{\flat,\mu} = p^\mu - \frac{m^2}{2p.n} n^\mu,
  \label{eq:pflatdef}
\end{equation}
where $p^{\mu}$ is the momentum of an on-shell massive fermion with $p^2=m^2$. The momenta $p^\flat$
and $n$ are both massless. The set of $u$ and $v$ spinors for the massive fermion can then be constructed
using the Weyl spinors for the massless momenta,
\begin{align}
  u_+(p,m) &= \frac{(\slashed{p}+m)|n\ra}{\spA{\fl{p}}{n}}, &
  u_-(p,m) &= \frac{(\slashed{p}+m)|n]}{\spB{\fl{p}}{n}},
  \label{eq:massivespinors1}  \\
  v_-(p,m) &= \frac{(\slashed{p}-m)|n\ra}{\spA{\fl{p}}{n}}, &
  v_+(p,m) &= \frac{(\slashed{p}-m)|n]}{\spB{\fl{p}}{n}}.
  \label{eq:massivespinors2}
\end{align}
The fact that the helicity state depends on the choice of reference vector means that positive and
negative helicities are no longer independent as they are in the massless case:
\begin{equation}
  u_-(p,m) = \frac{\spA{\fl{p}}{n}}{m} \left(u_+(p,m)\bigg|_{\fl{p} \leftrightarrow n}\right).
  \label{eq:uminus2plus}
\end{equation}
For a $t\bar{t}$ pair this means that we only need to compute one helicity configuration with
general reference vectors, say $++$, and we will obtain enough information for all four possible spin configurations.

We can then organise amplitudes for the $++$ configuration into a basis of spinor structures which
parametrise the dependence on the two reference vectors $n_1$ and $n_2$. There are four independent
terms in this basis, matching up with the four configurations in the complete system. For the
$L$-loop $\bar{t}tgg$ amplitude this is 
\begin{align}
  A^{(L)}(1_{\bar{t}}^+,2_t^+,3^{h_3},4^{h_4};n_1,n_2) = m\frac{\Phi^{h_3h_4}}{\spA{\fl{1}}{n_1}\spA{\fl{2}}{n_2}}
  \Bigg(
    \spA{n_1}{n_2}
    &A^{(L),[1]}(1_{\bar{t}}^+,2_t^+,3^{h_3},4^{h_4})\nonumber\\ +
    \frac{\spA{n_1}{3}\spA{n_2}{4}}{\spA34}
    &A^{(L),[2]}(1_{\bar{t}}^+,2_t^+,3^{h_3},4^{h_4})\nonumber\\ +
    \frac{s_{34} \spA{n_1}{3}\spA{n_2}{3}}{\spAA3{14}3}
    &A^{(L),[3]}(1_{\bar{t}}^+,2_t^+,3^{h_3},4^{h_4})\nonumber\\ +
    \frac{s_{34} \spA{n_1}{4}\spA{n_2}{4}}{\spAA4{13}4}
    &A^{(L),[4]}(1_{\bar{t}}^+,2_t^+,3^{h_3},4^{h_4})
  \Bigg)
  \label{eq:ttgg_spinbasis}
\end{align}
where $\Phi$ is a phase factor depending on the helicities of the gluons.
There are a few things to note about this expression:
\begin{enumerate}
  \item The sub-amplitudes $A^{(L),[a]}$ are defined to be dimensionless and free of the spinor phases.
  \item $A^{(L),[a]}$ depends only on two variables.
  \item The gluon phases for the $++$ and $+-$ configuration are taken to be
    \begin{align*}
      \Phi_{++} &= \frac{\spB34}{\spA34}, & \Phi_{+-} &= \frac{\spAB413}{\spAB314}.
      \label{eq:gluonphasefactors}
    \end{align*}
  \item The basis of spin structures has been chosen to prefer $\spA{n_1}{n_2}$ over
    $\spA{n_1}{4}\spA{n_2}{3}$. The sub-amplitudes appear to be simpler in this basis. 
\end{enumerate}

The computation of the sub-amplitudes can then be obtained from four different evaluations of the
full amplitude with a rational kinematic configuration with four choices of the reference vectors. These
evaluations can then be used to make a linear system which is solved to find the sub-amplitudes
$A^{(L),[a]}$. Explicitly,
\begin{align}
  A^{(L)}(3,3) &=
  m\,\Phi^{h_3h_4} \frac{s\spA34} {\spA{\fl{1}}{3}\spA{\fl{2}}{3} \spAB413} A^{(L),[4]},  \\
  A^{(L)}(4,4) &=
  -m\,\Phi^{h_3h_4} \frac{s\spA34} {\spA{\fl{1}}{4}\spA{\fl{2}}{4} \spAB314} A^{(L),[3]}, \\
  A^{(L)}(3,4) &=
  m\,\Phi^{h_3h_4} \frac{\spA34} {\spA{\fl{1}}{3}\spA{\fl{2}}{4}} A^{(L),[1]}, \\
  A^{(L)}(4,3) &=
  -m\,\Phi^{h_3h_4} \frac{\spA34} {\spA{\fl{1}}{4}\spA{\fl{2}}{3}} (A^{(L),[1]}+A^{(L),[2]}),
  \label{eq:projections}
\end{align}
where we have dropped the particle labels on the $A$ functions for simplicity.

\subsection{Generating a rational parametrisation of the kinematics}

We start from a rational configuration with six massless particles generated via momentum twistors
\cite{Hodges:2009hk,Badger:2013gxa}. This configuration will depend on eight independent parameters. The six particles
are then used to form a configuration with specific choices for the reference momenta. We label the six massless
momenta as $q_1,\dots,q_6$ and the resulting $t\bar{t}$ system as $p_1,p_2,p_3,p_4$ where,
\begin{align}
  p_1 &= q_1+q_2,&
  p_2 &= q_3+q_4,&
  p_3 &= q_5,&
  p_4 &= q_6,
\end{align}
\begin{align}
  q_1 \cdot q_2 &= q_3 \cdot q_4,&
  \spA{q_2 q_5} &= 0,&
  \spB{q_2 q_5} &= 0,&
  \spA{q_4 q_5} &= 0,&
  \spB{q_4 q_5} &= 0.
\end{align}
The ordering of the momenta $q_1,\dots,q_6$ is important to find a simple rational solution to the last four
constraints. From the on-shell momenta $p_1,\dots,p_4$ it is straightforward to generate rational
spinors for $p_1^\flat,p_2^\flat$ using four different choices of reference vectors: $(n_1,n_2) =
\{(3,3),(4,4),(3,4),(4,3)\}$. The variables of the original six-point configuration, i.e. those of
the $q_i$ momenta, are changed so that we can use conventional Mandelstam invariants,
\begin{align}
  & s = (p_3+p_4)^2 = 2p_3 \cdot p_4, &
  & t = (p_2+p_3)^2 = 2p_2 \cdot p_3 + m_t^2.
\end{align}
This procedure would also work for high multiplicity amplitudes though a careful choice of variables
in the parametrisation may be necessary to obtain manageable algebraic complexity.

The final results for the sub-amplitudes in Eq.~\eqref{eq:ttgg_spinbasis} will be expressed using kinematic variables $x$ and $y$ defined by 
\begin{equation}\label{eq:xydef}
-\frac{s}{m_t^2}=\frac{(1-x)^2}{x}, \qquad \frac{t}{m_t^2}=y. 
\end{equation} 
This choice rationalises the square root,
\begin{equation}
\sqrt{s \; (s-4 m_t^2)}
\nonumber
\end{equation}
that appears in the master integral and amplitude computations. For the $\bar{t}tgg$ amplitudes that do not involve elliptic sectors,
the rationalisation of such a square root allows us to perform a computation within 
finite field arithmetic without the need to introduce additional variables. It also important to
make a rational change of variables when solving the differential equations of the master integrals.

\subsection{Tree-level sub-amplitudes}

Following the procedure above allows us to directly evaluate the relevant colour ordered Feynman
diagrams to obtain simple expressions at tree level:
\begin{align}
  A^{(0),[1]}(1_{\bar{t}}^+,2_t^+,3^+,4^+) 
  &= -\frac{1}{1-y}, \nonumber\\
  A^{(0),[2]}(1_{\bar{t}}^+,2_t^+,3^+,4^+) &= A^{(0),[3]}(1_{\bar{t}}^+,2_t^+,3^+,4^+) = A^{(0),[4]}(1_{\bar{t}}^+,2_t^+,3^+,4^+)
  = 0,
\end{align}
and
\begin{align}
  A^{(0),[1]}(1_{\bar{t}}^+,2_t^+,3^+,4^-) &= A^{(0),[3]}(1_{\bar{t}}^+,2_t^+,3^+,4^-)
  = \frac{(x-y)(1-xy)}{(1-x)^2(1-y)}, \nonumber\\
  A^{(0),[2]}(1_{\bar{t}}^+,2_t^+,3^+,4^-) &= A^{(0),[4]}(1_{\bar{t}}^+,2_t^+,3^+,4^-)
  = 0.
\end{align}
The procedure for generating loop level expressions does not change although the expressions require,
as usual, tensor integral reduction to be performed. Our implementation of these steps is given in
the subject of the following sections. Our final aim is to present the finite remainders of the sub-amplitudes after
the subtraction of infrared and ultraviolet poles.

\section{Amplitude reduction}

The reduction of the helicity amplitudes, generated using Feynman diagrams, is performed directly to
special functions using finite field arithmetic in the \textsc{FiniteFlow} framework~\cite{Peraro:2019svx}. Recent years
have seen a growth in the popularity of loop amplitude computations with finite field arithmetic for
problems with many external scales where the algebraic complexity is too high for conventional
approaches. There are however many potential bottlenecks in amplitude reduction and it has been
shown to be important to control as many steps in the computation as possible. Amplitudes requiring
dimensional regularisation can cause unnecessary large intermediate steps since information that
vanishes in four dimensions is retained. Unitarity cut based approaches~\cite{Abreu:2020xvt} and specially designed
tensor projectors~\cite{Chen:2019wyb,Peraro:2019cjj,Peraro:2020sfm} have been used successfully in recent high multiplicity loop amplitude
computations in massless theories.

In our setup the diagram generation, colour ordering and spinor-helicity algebra are performed with the help of
\textsc{Qgraf}~\cite{Nogueira:1991ex}, \textsc{Form}~\cite{Kuipers:2012rf,Ruijl:2017dtg},
\textsc{Spinney}~\cite{Cullen:2010jv} and \textsc{Mathematica} scripts. During this phase we identify a set of topologies that are
independent at the integrand level. The numerators of these topologies are then separated into loop momentum dependent structures and
coefficients depending on the external kinematics. The latter are then evaluated using the rational
parametrisation described in the previous section. Up to this point we follow a strategy
described in, for example, Ref.~\cite{Hartanto:2019uvl}. The only difference is that the internal masses
must also be tracked.

The result of these steps is not yet suitable for processing with integration-by-parts identities
since the loop dependent numerators must be transformed into a basis of independent scalar products
thereby upgrading the topologies to complete families of Feynman integrals. In the case of
four-particle amplitudes an additional integration over the spurious space in the loop momenta must
also be performed. We employ a transverse integration approach similar to the one outlined in
Ref.~\cite{Mastrolia:2016dhn}.

For clarity, we present some additional details of these steps that are specific to our problem. An interested reader may wish to
refer to Refs.~\cite{Badger:2017jhb,Badger:2018enw,Badger:2021nhg} for a more complete discussion and other applications. All steps described from this point to the
reduction onto a basis of special functions have been constructed using \textsc{FiniteFlow} graphs.
After colour ordering and helicity amplitude processing the amplitude takes the following form,
\begin{equation}
A^{(L),h}(\lbrace p \rbrace) = \int \prod_{j=1}^{L} \frac{d^d k_j}{i \pi^{d/2} e^{-\eps \gamma_E}} \sum_{T}
\frac{N_T^h(d_s,\lbrace k \rbrace,\lbrace p \rbrace)}{\prod_{\alpha\in T} D_\alpha (\lbrace k \rbrace,\lbrace p \rbrace)},
\label{eq:calc_helicityamplitude}
\end{equation}
where $d = 4-2\eps$, $d_s = g_\mu^\mu$ and $k_i$ are the spacetime dimension, spin dimension and loop momenta, respectively.
$T$ is a set of distinct diagram topologies contributing to the colour-stripped helicity amplitude $A^{(L),h}(\lbrace p \rbrace)$.
A diagram topology may include contributions from more than one Feynman diagram.
To parameterise the loop integrand, we decompose the $d$-dimensional loop momenta $k_i$ into 
a four-dimensional part $\bar{k}_i$ and an extra-dimensional part $\tilde{k}_i$,
\begin{equation}
k_i^\mu = \bar{k}_i^\mu + \tilde{k}_i^\mu.
\end{equation}
Since the external momenta are in four dimensions, the extra dimensional part of the loop momenta
only appears in the numerator function as
$\mu_{ij} = -\tilde{k}_i \cdot \tilde{k}_j$. After substituting massive spinors in terms of
massless ones according to Eqs.~\eqref{eq:massivespinors1}~-~\eqref{eq:massivespinors2} and 
performing t'Hooft algebra, the numerator function contains the following loop-momentum dependent terms
\begin{equation}
k_i \cdot k_j, \;
\bar{k}_i \cdot p_j, \;
\mu_{ij}, \;
\spAB{p_a}{\bar{k}_i}{p_b}, \;
\spAA{p_a}{\bar{k}_i\bar{k}_j}{p_b}, \;
\spBB{p_a}{\bar{k}_i\bar{k}_j}{p_b},
\label{eq:k-dependent-term}
\end{equation}
where momenta $p_a$ and $p_b$ are massless and $p_j$ can either be massive or massless. As indicated
above, to apply IBP reduction we must first express these loop momentum structures in a basis of the
propagators $D_\alpha$ and a set of irreducible scalar products (ISPs) which define a family of
Feynman integrals.

\begin{figure}[t]
  \begin{center}
    \includegraphics[width=0.65\textwidth]{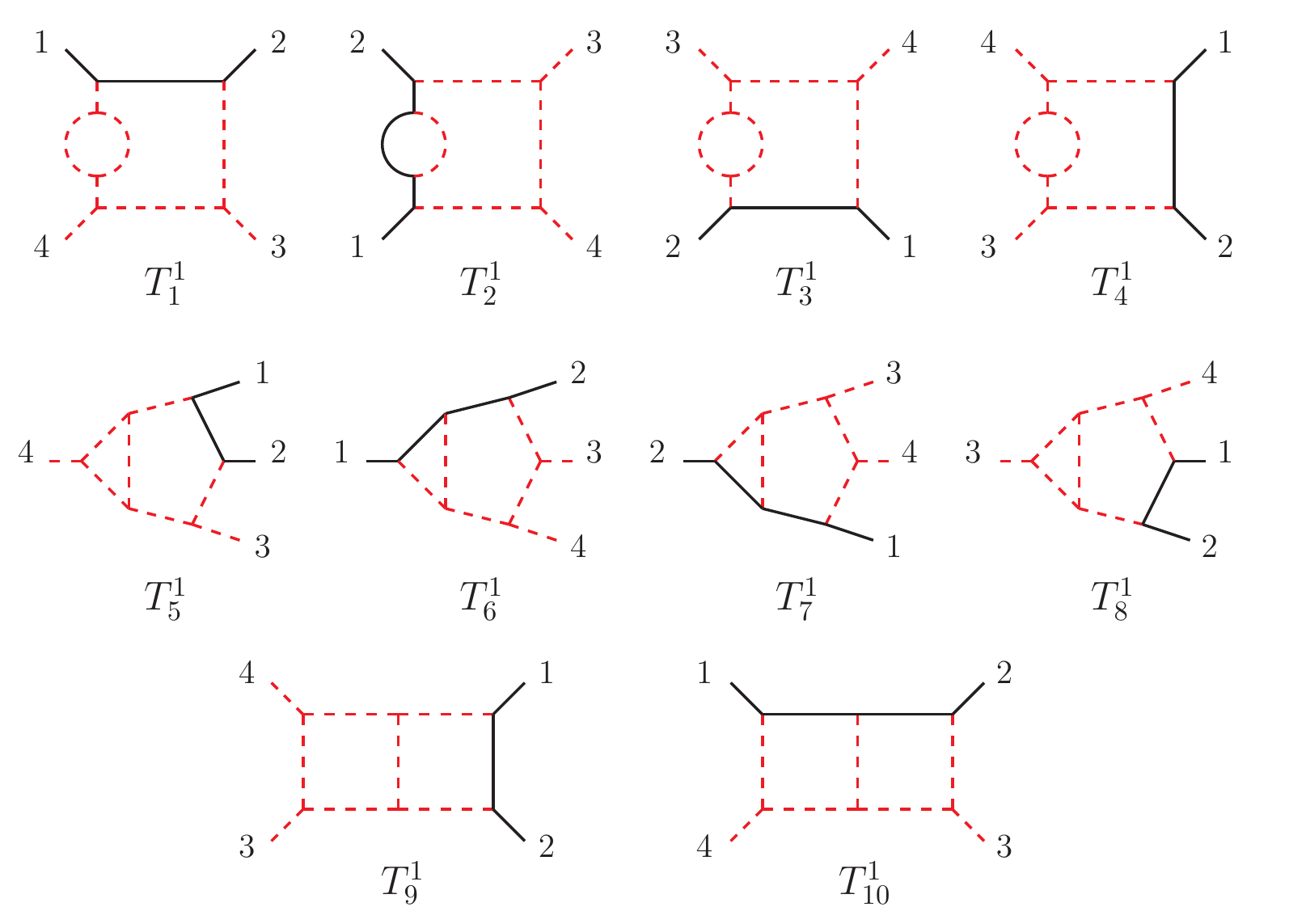} 
  \end{center}
  \caption{Maximal topologies for the $N_c^2$ part of planar two-loop $\bar{t}tgg$ amplitude.
  Black-solid lines represent top quarks, red-dashed lines represent gluons.}
  \label{fig:maxtopo_ncp2}
\end{figure}

We make sure that we introduce the minimal number of integral families by mapping all distinct
diagram topologies, $T$, to a set of master topologies which have the maximum number of
propagators. The master topologies for the $N_c^2$ part of the two-loop planar $\bar{t}tgg$
amplitude are shown in  Fig.~\ref{fig:maxtopo_ncp2}. Each topology $T$ will map to at least one
master topology. The numerators are then expressed in terms of the loop momenta of the master
topology. We split the four-dimensional loop momenta $k_i$ into physical parts
$\bar{k}_{\parallel,i}^\mu$ and spurious parts $\bar{k}_{\perp,i}^\mu$, 
\begin{equation}
\bar{k}_i^\mu = \bar{k}_{\parallel,i}^\mu + \bar{k}_{\perp,i}^\mu,
\end{equation}
and we further expand them into a physical ($v^\mu$) and a spurious ($w^\mu$) spanning bases
\begin{equation}
\bar{k}_{\parallel,i}^\mu = \sum_{j=1}^{d_\parallel} a_{ij} v_j^\mu, \qquad \qquad \bar{k}_{\perp,i}^\mu = \sum_{j=1}^{d_\perp} b_{ij} w_j^\mu \,.
\label{eq:loopmomexpansion}
\end{equation}
$d_\parallel = \mathrm{min}(4,n_\mathrm{ext})$ is the dimension of the $4d$ physical space, $n_\mathrm{ext}$
is the number of independent momenta of a given topology and $d_\perp = 4-d_\parallel$ is the dimension of the $4d$ spurious space.
In addition, the physical vectors are orthogonal to the spurious vectors, $v_i \cdot w_j = 0$.
The coefficients of the physical spanning basis are functions of inverse propagators and physical ISPs,
$a_{ij} = a_{ij}(D_\alpha,\mathrm{ISPs})$, while the coefficients of the spurious spanning basis are functions
of spurious ISPs and extra dimensional ISPs, $b_{ij} = b_{ij}(k_i \cdot w_j,\mu_{ij})$.
$a_{ij}$ and $b_{ij}$ are determined by solving a linear system of equations obtained by contracting 
Eq.~\eqref{eq:loopmomexpansion} with $v_i$ and $w_i$. For the $\bar{t}tgg$ amplitudes, all the
maximal topologies have four external momenta, hence $n_\mathrm{ext} = 3$, 
$d_\parallel = 3$ and $d_\perp = 1$, and for instance, we can choose the following set of spanning vectors
\begin{equation}
v^\mu = \lbrace p_1^\mu, p_3^\mu, p_4^\mu \rbrace, \qquad w^\mu = \lbrace \omega^\mu \rbrace.
\end{equation}
Explicitly, we used:
\begin{equation}
\omega^\mu = \frac{1}{2} \big(\spAXB{3}{4}{1}{\sigma^\mu}{3}-\spAXB{3}{\sigma^\mu}{1}{4}{3} \big),
\end{equation}
though any vector satisfying $\omega\cdot p_1 = \omega\cdot p_3 = \omega\cdot p_4 = 0$ will suffice. 
Numerator terms with odd powers of $k_i \cdot w_j$ vanish under loop integration, while the terms
with even powers can be cast into linear combination of propagator denominators $D_\alpha$ and physical ISPs.
For example,
\begin{equation}
k_i \cdot w_k \; k_j\cdot w_k = \frac{w_k^2}{d_\perp - 2\eps} \left( k_i\cdot k_j - k_{\parallel,i}\cdot k_{\parallel,j} \right),
\end{equation}
where $k_i\cdot k_j$ can be directly written in terms of $D_\alpha$ while $k_{\parallel,i}\cdot
k_{\parallel,j}$ can be expressed in terms of $D_\alpha$ and physical ISPs according to
Eq.~\eqref{eq:loopmomexpansion}.  Similarly for the extra dimensional ISP,
\begin{equation}
\mu_{ij} = \frac{-2\eps}{d_\perp - 2\eps} \left( k_i\cdot k_j - k_{\parallel,i}\cdot k_{\parallel,j} \right).
\end{equation}
After the replacements have been made the amplitude is recast into a form suitable for IBP
reduction,
\begin{equation}
A^{(L),h}(\lbrace p \rbrace)  = \sum_{T} \sum_{i} c^h_{T,i}(\eps,\lbrace p \rbrace) \; G_{T,i}(\eps, \lbrace p \rbrace).
\label{eq:calc_ampint1} 
\end{equation}
Note that there may be overlap in sets of Feynman integrals $G_{T,i}(\lbrace k \rbrace, \lbrace p \rbrace)$ appearing in each numerator topology $T$.
This is in contrast to the integrand reduction approach taken in~\cite{Hartanto:2019uvl} where there is no overlap in $G_{T,i}(\lbrace k \rbrace, \lbrace p \rbrace)$
between numerator topologies. Combining contributions from all numerator topologies we have
\begin{equation}
A^{(L),h}(\lbrace p \rbrace)  = \sum_{k} \tilde{c}^h_{k}(\eps,\lbrace p \rbrace) \; \tilde{G}_{k}(\eps, \lbrace p \rbrace),
\label{eq:calc_ampint2}
\end{equation}
where $\tilde{G}_{k}$ is a unique set of Feynman integrals obtained from all the $G_{T,k}$ in Eq.~\eqref{eq:calc_ampint1}.
Before proceeding with IBP reduction, at this stage we combine the bare helicity amplitude with the mass renormalisation counterterm contributions
according to Eqs.~\eqref{eq:mren_1loop}~and~\eqref{eq:mren_2loop}, for the one- and two-loop amplitudes, to obtain a gauge invariant, mass renormalised amplitude,
\begin{align}
A_{\mren}^{(L),h}(\lbrace p \rbrace)  & = A^{(L),h}(\lbrace p \rbrace) + A_{\mct}^{(L),h}(\lbrace p \rbrace), \\
& = \sum_{k} c^{\mren,h}_{k}(\eps,\lbrace p \rbrace) \; \tilde{G}_{k}(\eps, \lbrace p \rbrace).
\label{eq:calc_ampmren} 
\end{align}
The mass renormalisation counterterms appearing in Eqs.~\eqref{eq:mren_1loop}~and~\eqref{eq:mren_2loop} have to be written in the integral representation,
in order to be combined with the bare amplitude.

The integrals appearing in Eq.~\eqref{eq:calc_ampmren} are further reduced to a set of master integrals via IBP identities. 
The IBP relations are generated in \textsc{Mathematica} with the help of \textsc{LiteRed}~\cite{Lee:2012cn}, 
and solved numerically over finite fields within the \textsc{FiniteFlow} framework~\cite{Peraro:2019svx}
using the Laporta algorithm~\cite{Laporta:2001dd}. The mass renormalised helicity amplitude is now written in a basis of master integrals $\mi_k$,
\begin{equation}
A_{\mren}^{(L),h}(\lbrace p \rbrace)  = \sum_{k} c^{\mathrm{IBP},h}_{k}(\eps,\lbrace p \rbrace) \; \mi_{k}(\eps, \lbrace p \rbrace).
\label{eq:calc_ampMI} 
\end{equation}
Let us note that not all of the maximal topologies that appear in an amplitude are processed in the IBP system, as some of them can be mapped
into the other maximal topologies. For example, maximal topologies $T^1_1,\dots,T^1_4$ in Fig.~\ref{fig:maxtopo_ncp2} can be mapped to 
topologies $T^1_5,\dots,T^1_8$ by choosing an appropriate set of physical ISPs (auxiliary propagators) for each of those topologies.
The master integral representation can be used to check Ward identities on the mass-renormalised amplitude by replacing
gluon polarisation with its momentum in the numerator construction and reconstructing Eq.~\eqref{eq:calc_ampMI} over a prime field
to check if it indeed vanishes.

If the analytic solutions of the master integrals appearing in Eq.~\eqref{eq:calc_ampMI} 
for the planar $\bar{t}tgg$ amplitudes are available, we can proceed further by writing the master integrals 
as a linear combination of special function monomials $m\left(f(\lbrace p \rbrace)\right)$ and perform Laurent expansions,
\begin{equation}
A_{\mren}^{(L),h}(\lbrace p \rbrace)  = \sum_{k} \sum_{l=n(L)}^{0} \eps^l \; c^{\mathrm{exp},h}_{kl}(\lbrace p \rbrace) \; m_k\left(f(\lbrace p \rbrace) \right)
+ \cO(\eps),
\label{eq:ampexpanded}
\end{equation}
where $n(L)$ is the power of the deepest pole that can appear in the $L$-loop amplitude. For one- and two-loop cases, $n(1)=-2$ and $n(2)=-4$.
To obtain the simplest representation of the amplitude we further subtract the UV and IR poles from the mass-renormalised amplitude.
We first need to write the UV and IR poles, given in Eqs.~\eqref{eq:ampren_1loop},~\eqref{eq:ampren_2loop},~\eqref{eq:amprenorm1loop}~and~\eqref{eq:amprenorm2loop}
for the case of planar $\bar{t}tgg$ one- and two-loop amplitudes,
in the same special function monomial basis as in Eq.~\eqref{eq:ampexpanded},
\begin{align}
P_{\mathrm{UV}}^{(L),h}(\lbrace p \rbrace)  
  & = \sum_{k} \sum_{l=n(L)}^{0} \eps^l \; c^{\mathrm{UV},h}_{kl}(\lbrace p \rbrace) \; m_k\left(f(\lbrace p \rbrace) \right) + \cO(\eps), 
\label{eq:poleexpandedUV}
\\
P_{\mathrm{IR}}^{(L),h}(\lbrace p \rbrace)  
  & = \sum_{k} \sum_{l=n(L)}^{0} \eps^l \; c^{\mathrm{IR},h}_{kl}(\lbrace p \rbrace) \; m_k\left(f(\lbrace p \rbrace) \right) + \cO(\eps),
\label{eq:poleexpandedIR}
\end{align}
and subtract them from the mass-renormalised amplitude to obtain the finite remainder,
\begin{align}
F^{(L),h}(\lbrace p \rbrace) & =  A_{\mren}^{(L),h}(\lbrace p \rbrace) 
+ P_{\mathrm{UV}}^{(L),h}(\lbrace p \rbrace)
- P_{\mathrm{IR}}^{(L),h}(\lbrace p \rbrace)
\label{eq:finiteremainder1} \\
 & = \sum_{k} \; c^{\mathrm{F},h}_{k}(\lbrace p \rbrace) \; m_k\left(f(\lbrace p \rbrace) \right) + \cO(\eps).
\label{eq:finiteremainder2}
\end{align}
The coefficients $c^{\mathrm{F},h}_{k}(\lbrace p \rbrace)$
appearing in~\eqref{eq:finiteremainder2}, however,
are not all independent.  We exploit this fact in order to simplify
both the result and the reconstruction of its analytic expression as
follows.  First, we sort all the coefficients by their complexity,
which is estimated from their total degree.  The total degree, in
turn, can be quickly determined via a univariate fit, as explained in~\cite{Peraro:2016wsq}.  
We then find vanishing linear
combinations of these coefficients by solving the linear fit problem
\begin{equation}
  \label{eq:linearrels}
  \sum_k y_k\, c^{\mathrm{F},h}_{k}(\lbrace p \rbrace) = 0
\end{equation}
with respect to the unknowns $y_k$, which are rational numbers.
This allows us to find linear relations between the coefficients, which
rewrite the more complex ones in terms of simpler ones.  After
applying the linear relations between coefficients of special
function monomials, we arrive at
\begin{equation}
F^{(L),h}(\lbrace p \rbrace)  = \sum_{k} \; \bar{c}^{\mathrm{F},h}_{k}(\lbrace p \rbrace) \; \bar{m}_k\left(f(\lbrace p \rbrace) \right) + \cO(\eps),
\label{eq:finiteremainderindep}
\end{equation}
where $\bar{c}^{\mathrm{F},h}_{k}$ are the independent coefficients of the new
special function monomials $\bar{m}_{k}(f)$, which are linear combinations of the monomials
appearing in Eq.~\eqref{eq:finiteremainder2}.  Therefore, functional
reconstruction only needs to be applied to the independent
coefficients $\bar{c}^{\mathrm{F},h}_{k}$.  This yields a significantly
simpler result than the one in Eq.~\eqref{eq:finiteremainder2}, and also
reduces the number of evaluations needed for its reconstruction.

Having set up the numerical algorithm suitable for computation over finite fields, 
starting from the Feynman diagram numerators, to evaluate $\bar{c}^{\mathrm{F},h}_{k}$
in the $L$-loop finite remainder $F_{n}^{(L),h}(\lbrace p \rbrace)$ using the \textsc{FiniteFlow} framework,
the analytic forms of $\bar{c}^{\mathrm{F},h}_{k}$ are obtained by using the
multivariate reconstruction method described in~\cite{Peraro:2016wsq}, after performing several numerical evaluations.

\section{Master integrals}
\label{sec:masterintegrals}

To derive an analytic representation of the helicity amplitudes, particularly when we are interested in computing the finite remainder,
analytic expressions of master integrals appearing in the one- and two-loop amplitudes are required.
The solutions of all one- and two-loop master integrals appearing in the $A^{(2),1}$, $A^{(2),N_l}$, $A^{(2),N_l^2}$, $A^{(2),N_l N_h}$ 
and $A^{(2),N_h^2}$ amplitudes in Eq.~\eqref{eq:flamplitude2loop} can be expressed in terms of 
multiple polylogarithms (MPLs)~\cite{goncharov2011multiple,Moch_2002,Borwein:1999js} and have been completed recently~\cite{Bonciani:2010mn,Mastrolia:2017pfy}.
For the two-loop amplitude $A^{(2),N_h}$ involving a single top-quark closed-loop, 
the master integrals also involve elliptic generalisations of MPLs~\cite{Adams:2018bsn,Adams:2018kez}. Before we proceed further, it is useful to introduce the notion of a sector (or a topology). A sector is defined by the set of propagators with positive exponents.

\begin{figure}[t]
  \begin{center}
    \includegraphics[width=0.75\textwidth]{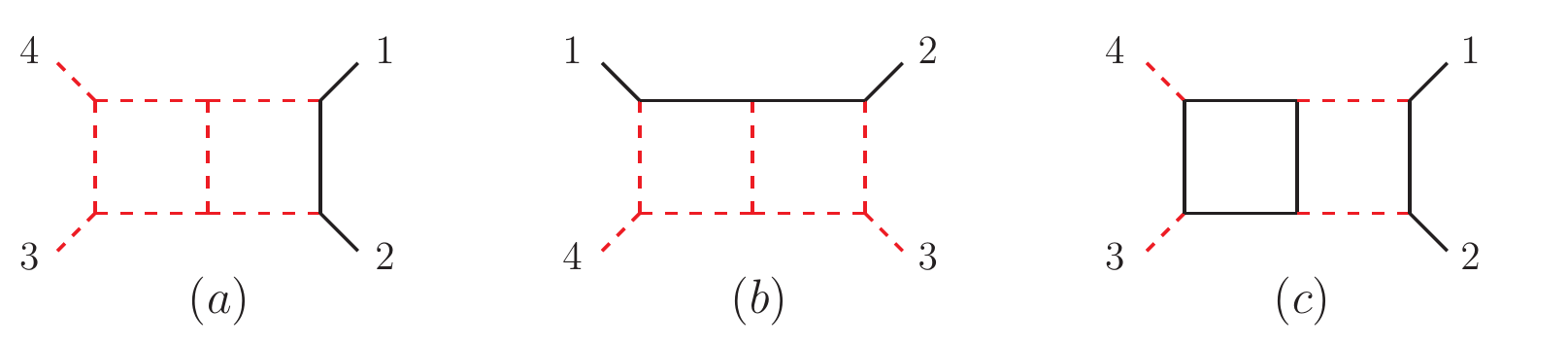} 
  \end{center}
  \caption{Master integral topologies with maximal number of propagators (7) that appear in the leading colour 2-loop amplitude for $\ttgg$. 
  Black-solid lines represent massive particles, red-dashed lines represent massless particles.}
  \label{fig:maxtopo_mis}
\end{figure}

\begin{figure}[t]
  \begin{center}
    \includegraphics[width=0.45\textwidth]{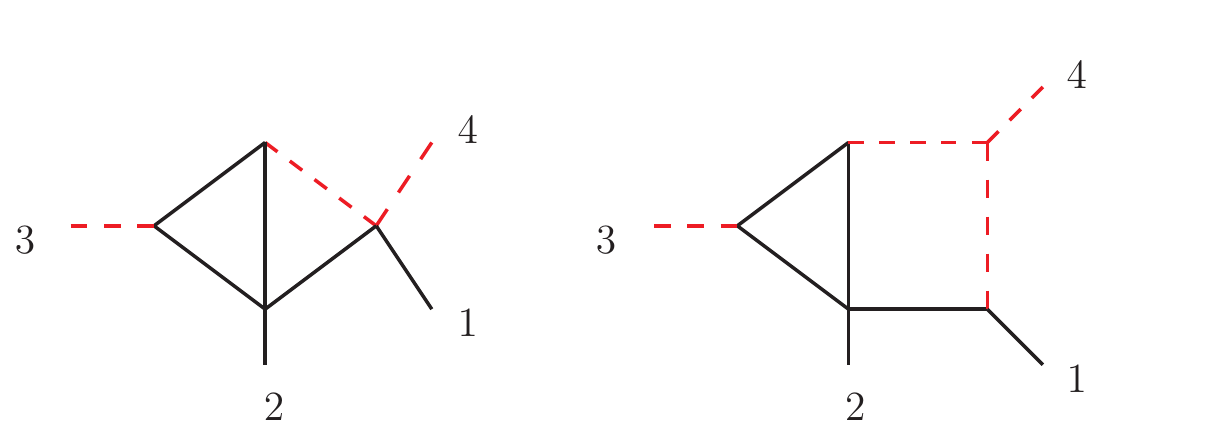} 
  \end{center}
  \caption{New master integral topologies appearing in the amplitude with a closed top-quark loop $A^{(2),N_h}$. 
  Black-solid lines represent massive particles, red-dashed lines represent massless particles.}
  \label{fig:new_mis}
\end{figure}

In our work, we employ the following master integrals to obtain analytic helicity amplitudes:
\begin{itemize}
\item \underline{$A^{(2),1}$, $A^{(2),N_l}$, $A^{(2),N_l^2}$} :
There are 36 master integrals appearing in the $A^{(2),1}$ amplitude. 
We have used the results in~\cite{Mastrolia:2017pfy} for the relevant master integrals. 
The basis of master integrals involves the topologies with 7 propagators (double-box) shown 
in Fig.~\ref{fig:maxtopo_mis} (topology $a$ and $b$).  The master integrals required for the 
computation of  $A^{(2),N_l}$ and $A^{(2),N_l^2}$ can be obtained from a subset 
of the master integral basis of $A^{(2),1}$.

\item \underline{$A^{(2),N_h}$}:
There are 55 master integrals in the amplitude with a single top-quark closed loop. 44 of them belong to the \textit{topbox} family, i.e. 
the planar double-box integral family contributing to $A^{(2),N_h}$ (topology $c$ in Fig.~\ref{fig:maxtopo_mis}), 
computed in~\cite{Adams:2018kez}. 
Results for the 9 master integrals that are not part of the topbox family are obtained from~\cite{Mastrolia:2017pfy}.
Analytic expressions for the remaining 2 master integrals (shown in Fig.~\ref{fig:new_mis}), to the best of our knowledge, are not available in the literature.
In this work we have computed those two integrals using the method of differential equations. 
The details of the computation will be covered in the next subsection.

\item \underline{$A^{(2),N_l N_h}$ and $A^{(2),N_h^2}$}:  Both amplitudes are of one-loop squared type, 
and the master integrals needed are the one-loop bubbles and triangle up to weight 4, 
which are already available from the one-loop amplitudes computation up to $\cO(\eps^2)$.

\end{itemize}

\subsection{New master integrals}
\begin{figure}[h]
  \begin{center}
    \includegraphics[width=0.35\textwidth]{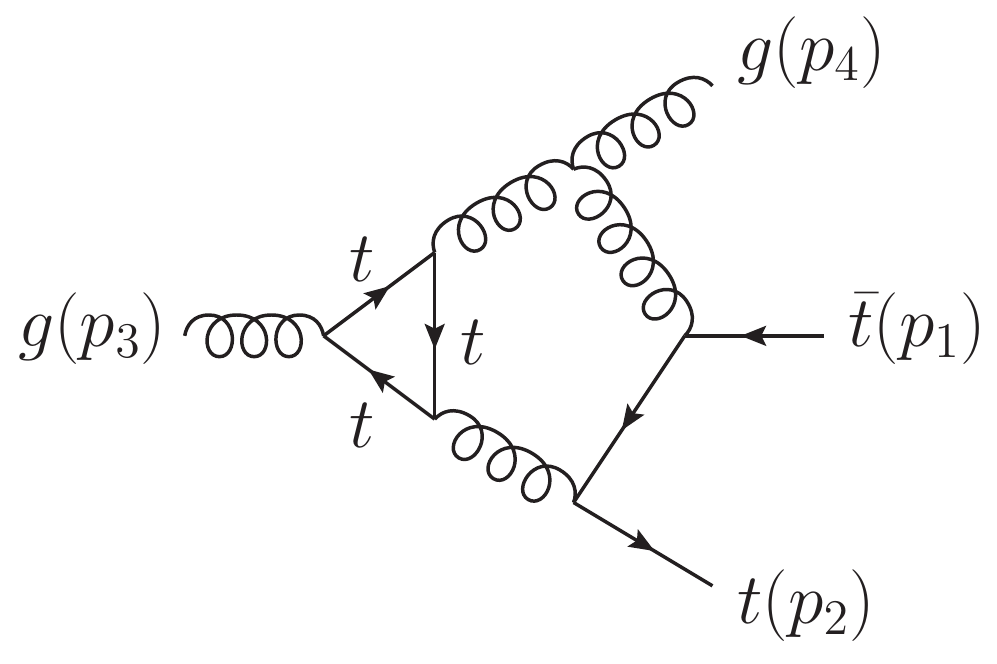} 
  \end{center}
  \caption{Feynman diagram contributing to $A^{(2),N_h}$ that leads to the new master integral topologies shown in Fig.~\ref{fig:new_mis}.
  The solid lines denoted the top quark whereas spiral lines denote the gluons. All external momenta are on-shell.}
  \label{fig:t8pentatriangle}
\end{figure}
\begin{figure}[h]
  \begin{center}
    \includegraphics[width=0.8\textwidth]{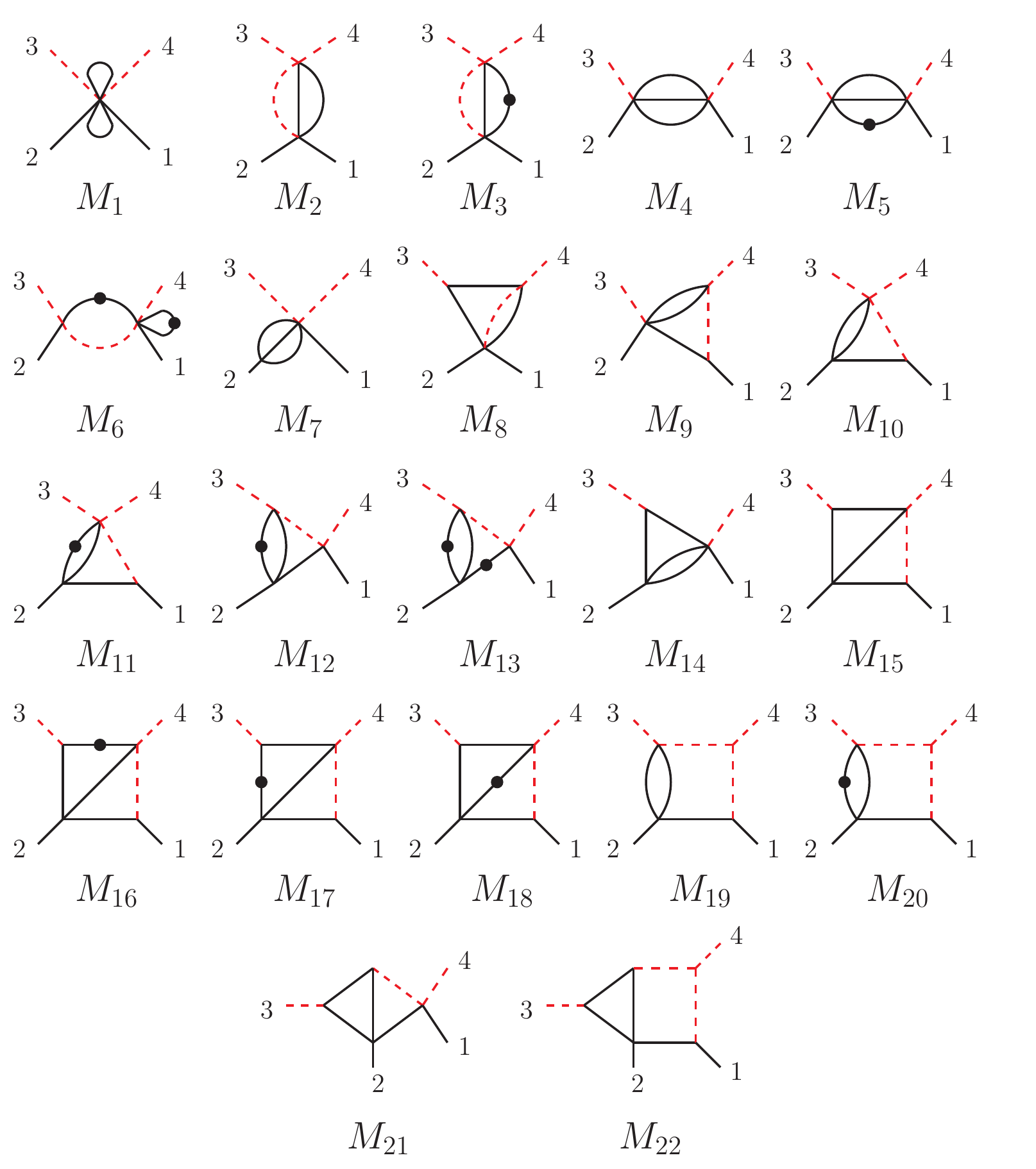} \\
  \end{center}
  \caption{Master integral basis for the pentatriangle topology shown in Fig.~\ref{fig:t8pentatriangle}.
  Black-solid lines represent massive particles, red-dashed lines represent massless particles.}
  \label{fig:t8masters}
\end{figure}

The new master integral topologies that appear in $A^{(2),N_h}$ arise from the Feynman diagram shown in Fig.~\ref{fig:t8pentatriangle}. The integral family for this penta-triangle topology is given by
\begin{equation}
\cI_{\nu_1 \nu_2 \nu_3 \nu_4 \nu_5 \nu_6 \nu_7 \nu_8 \nu_9} = 
e^{2\eps\gamma_E} \left(m_t^2\right)^{\nu-d}
\int \frac{d^d k_1}{i \pi^{d/2}} \frac{d^d k_2}{i \pi^{d/2}}  \;
\frac{D_8^{-\nu_8} D_9^{-\nu_9}}{D_1^{\nu_1} D_2^{\nu_2} D_3^{\nu_3} D_4^{\nu_4} D_5^{\nu_5} D_6^{\nu_6} D_7^{\nu_7} },
\end{equation}
where $\gamma_E$ denotes the Euler-Mascheroni constant, $\nu = \sum_{j=1}^{9} \nu_j$ and $d=4-2\eps$. The inverse propagators $D_i$ are given by 
\begin{align}
& D_1 = -k_1^2, && D_2 = -(k_1-p_4)^2, && D_3 = -(k_1+p_2+p_3)^2 + m_t^2, \nn
& D_4 = -(k_1+p_3)^2, && D_5 = -k_2^2 + m_t^2, && D_6 = -(k_2-p_3)^2 + m_t^2, \nn
& D_7 = -(k_1+k_2)^2 + m_t^2, && D_8 = -(k_2+p_4)^2, && D_9 = -(k_2+p_1+p_4)^2. \nn
\end{align}
We choose a basis of master integrals for which the irreducible scalar products corresponding to $D_8$ and $D_9$ are absent. 
Therefore we may label these master integrals by $\cI_{\nu_1 \nu_2 \nu_3 \nu_4 \nu_5 \nu_6 \nu_7}$. 
Using IBP identities, we obtain 22 master integrals, $\{ M_i \}_{i=1}^{22} $ which are shown in Fig.~\ref{fig:t8masters}.
We refer to this basis of master integrals as the pre-canonical basis, and they are given by
\begin{align}
 &   M_1 = \cI_{000001100},  &&    M_2 = \cI_{010001100}, &&    M_3 = \cI_{010002100}, \nn
 &   M_4 = \cI_{001010100},  &&    M_5 = \cI_{002010100}, &&    M_6 = \cI_{102020000}, \nn
 &   M_7 = \cI_{001001100},  &&    M_8 = \cI_{010011100}, &&    M_9 = \cI_{011010100}, \nn
 &M_{10} = \cI_{011001100},  && M_{11} = \cI_{011002100}, && M_{12} = \cI_{101001200}, \nn
 &M_{13} = \cI_{102001200},  && M_{14} = \cI_{001011100}, && M_{15} = \cI_{011011100}, \nn
 &M_{16} = \cI_{011021100},  && M_{17} = \cI_{011012100}, && M_{18} = \cI_{011011200}, \nn
 &M_{19} = \cI_{111001100},  && M_{20} = \cI_{111001200}, && M_{21} = \cI_{101011100}, \nn
 &M_{22} = \cI_{111011100}.  &&  &&  
\end{align}
A set of 15 integrals were identified as master integrals from the topbox family and were 
already computed in~\cite{Adams:2018kez}. They are
\begin{align}
 & M_1 = I_{1001000}, && M_2 = I_{0011100}, && M_3 = I_{0021100}, && M_4 = I_{1001001}, \nn
 & M_5 = I_{2001001}, && M_7 = I_{0101001}, && M_8 = I_{1011100}, && M_9 = I_{1001101}, \nn
 & M_{10} = I_{0011101}, && M_{11} = I_{0021101}, && M_{14} = I_{1101001}, && M_{15} = I_{1011101}, \nn
 & M_{16} = I_{2011101}, && M_{17} = I_{1021101}, && M_{18} = I_{1012101},  
\end{align}
where $I_{\nu_1 \nu_2 \nu_3 \nu_4 \nu_5 \nu_6 \nu_7}$ are the precanonical master integrals defined in 
Eq.~(8) of~\cite{Adams:2018kez}. 
A second set of 5 master integrals is available from~\cite{Mastrolia:2017pfy}, and they are identified as
\begin{align}
 & M_6 = \cT_{10}, && M_{12} = \cT_{21}, && M_{13} = \cT_{22}, && M_{19} = \cT_{31}, && M_{20} = \cT_{32},
 \label{eq:pentaintegrals}
\end{align}
where $\cT_i$ are master integrals defined in Sec.~6.2 of~\cite{Mastrolia:2017pfy}.   $M_{21}$ and $M_{22}$ are the missing integrals.
Note that these two integrals also appear in a planar double-box family that contributes to the subleading colour $\bar{t}tgg$ amplitude without 
closed fermion loops (topology $P_5$ in Figure~1 of~\cite{Chen:2019zoy}).

Apart from the $x$ and $y$ defined in Eq.~\eqref{eq:xydef}, it can be useful to use another coordinate system $\tilde{x}$ and $\tilde{y}$ defined as
\begin{equation}
-\frac{s}{m_t^2}=\frac{(1+\tilde{x}^2)^2}{\tilde{x}\; (1- \tilde{x}^2)}, \qquad y=1-\tilde{y} \,, 
\label{eq:xytilde}
\end{equation}
in order to rationalize some of the square roots appearing in the differential equations.

In the following we aim to bring all the master integrals in a canonical form \cite{Henn:2013pwa,Adams_2018}. 
For the master integrals not belonging to the topbox family we change the basis such that
\begin{equation}
d \vec{\cJ}\;=\; \eps A \vec{\cJ}, \quad A = A_x \; dx + A_y \; dy,
\end{equation}
where $A$ does not depend on $\eps$ and is rational in the kinematic variables $(x,y)$. In particular, the canonical basis for integrals in Eq.~\eqref{eq:pentaintegrals} along with the missing integrals from that family is chosen as
\begin{align}
\cJ_{6}  & =  -\eps^2 y M_{6}, \nn
\cJ_{12} & =  \eps^3 (1-y) M_{12}, \nn
\cJ_{13} & =  -\frac{3}{128} \eps (1-\eps)^2 (3-11\eps) (1-y) M_{1} \nn
         & \quad \;    -\frac{3}{128} \eps (1-2\eps) (2-3\eps) (1-3\eps) (1-y) M_{7} \nn
         & \quad \;  +\frac{1}{2} \eps^2 (1-y) (1+y) M_{13}, \\
\cJ_{19} & = -\eps^3 (1-2\eps) (1-y) M_{19} - 2 \eps^3 \frac{(1-y)(1+x)}{x} M_{20}, \nn
\cJ_{20} & = \eps^3 \frac{(1-y)(1-x)(1+x)}{2x}M_{20}, \nn
\cJ_{21} & = -\eps^4 (1-y) M_{21}, \nn
\cJ_{22} & = -\eps^4  \frac{(1-x)^2 (1-y)}{x} M_{22}. \nonumber
\end{align}
For the set of master integrals belonging to the topbox topology 
we follow the choice given in Sec.~6 of ~\cite{Adams:2018kez}. 
Here the form of the differential equation is slightly relaxed and the differential equation for the basis $\cJ$ has the form
\begin{equation}\label{eq:DEQ}
d \vec{\cJ} = A_{x,y}^{(0)} + \eps A_{x,y}^{(1)} \; \vec\cJ, 
\end{equation}
where $A^{(0)}$ is strictly lower triangular and $A^{(0)}$ and $A^{(1)}$ are independent of $\eps$.
$A_{x,y}^{(i)}$ is rational in the kinematic variables $(x,y)$, periods of elliptic curves and their derivatives \cite{Adams_2018}.
By considering them as independent variables, we derived the system of differential equations efficiently 
with the finite field sampling method using the \textsc{FiniteFlow} package~\cite{Peraro:2019svx}. 
We note that in the computation of  the $A^{(2),N_h}$ finite remainder, the coefficients of the special function monomials,
$\bar{c}^{\mathrm{F},h}_{k}$ in Eq.~\eqref{eq:finiteremainderindep}, are also rational functions in 
 $(x,y)$, periods of elliptic curves and their derivatives.
The solution for master integrals in terms of iterated integrals are presented up to 4 orders in $\epsilon$:
\begin{equation}
\cJ_i = \sum_{k=0}^{4}\eps^k \; \cJ_i^{(k)} + \cO(\eps^5). \label{eq:iter_J}
\end{equation}

\subsection{Iterated integrals}

Now that we have identified the necessary master integrals, let us discuss the mathematical objects
that can used to describe them. By iteratively solving the differential equation~\eqref{eq:DEQ}, we naturally obtain a representation of master integrals in terms of iterated integrals. Let us therefore start by reviewing the definition of Chen's iterated integrals \cite{Chen:1977oja}. Consider a path $\gamma$ on an $n$-dimensional manifold $M$ with starting point $x_i=\gamma(0)$ and end point $x_f=\gamma(1)$,
\begin{align*}
\gamma\; : \; [0,1] \rightarrow M.
\end{align*}
Let us also consider a set of differential 1-forms $\lbrace\omega_i \rbrace$ and their pullbacks to the interval $[0,1]$ which we denote by 
\begin{align}\label{eq:pullback}
f_j (\lambda)\;  d \lambda = \gamma^{*} \omega_j.
\end{align}
Then we define the $k$-fold iterated integral over $\omega_1,...,\omega_k$ along $\gamma$ as
\begin{align}\label{eq:IIDef}
I_\gamma (\omega_1,...,\omega_k ; \lambda)\; &= \; \int_0^\lambda d \lambda_1 f_1 (\lambda_1) \int_0^{\lambda_1} d \lambda_2 f_2 (\lambda_2)\; ... \int_0^{\lambda_{k-1}} d \lambda_k f_k (\lambda_k) \\
&= \; \int_0^\lambda d \lambda_1 f_1 (\lambda_1) \, I_\gamma (\omega_2,...,\omega_k ; \lambda_1) \,.
\end{align}
From the above definition it is clear that the derivative of an iterated integral is given by 
\begin{align*}
\frac{d}{d \lambda} I_\gamma (\omega_1,\omega_2,...,\omega_k ; \lambda)\; =\; f_1  (\lambda) \;I_\gamma (\omega_2,...,\omega_k;\lambda) \,.
\end{align*}
It can be shown that the iterated integrals defined in Eq.~\eqref{eq:IIDef} have various interesting properties~\cite{Brown:2011}. The most useful property for the purpose of this paper is concerning the decomposition of the path $\gamma$. Let $\alpha, \beta$: $[0,1] \rightarrow M$ be two paths such that $\beta(0) =\alpha(1) $ and let $\gamma = \alpha \beta$ be the path obtained by concatenating $\alpha$ and $\beta$. Then we can decompose
\begin{align}\label{eq:path_decomp}
I_\gamma (\omega_1,...,\omega_k ; \lambda)\; = \sum_{i=0}^n I_\beta(\omega_1,...,\omega_i ; \lambda)  \, I_\alpha(\omega_{i+1},...,\omega_n ; \lambda) \,,
\end{align}
where we define the $0$-fold integrals as 
\begin{align}
I_\gamma(; \lambda)\; = \;1.
\end{align}

Now that we are familiar with the notion of iterated integrals, let us consider the well-understood multiple polylogarithms (MPLs). MPLs are a special class of iterated integrals where the $1$-forms $\omega_i$ are such that
\begin{align}
\gamma^{*} \omega_j = \frac{d\lambda}{\lambda - c} \,,
\end{align}
for some $c \in \mathbb{C}$. Then we have
\begin{align}\label{eq:recursive}
G(z_1,...,z_k;y) = \int_0^y \frac{dy_1}{y_1 - z_1} G(z_2,...,z_k;y_1) \,.
\end{align}
Note that the recursion defined in~\eqref{eq:recursive} diverges whenever $z_k = 0$. We can regularise this divergence by setting 
\begin{align}
G(\underbrace{0,\dots,0}_{k \text{ times}};y) = \frac{1}{k!} \left( \ln y \right)^k.
\end{align}

The class of MPLs is very well understood and many tools for their algebraic manipulation and their numerical evaluation are available~\cite{Duhr:2019tlz,Panzer:2015ida,Bauer:2000cp}, it is however not always possible to express all master integrals in terms of these functions. Especially in computations involving massive internal particles we often encounter elliptic sectors which transcend the space of MPLs. In the following, we give a brief overview of the elliptic sectors we encounter in this work as well as the computational peculiarities that come along with them.

\subsection{Elliptic curves}
In this section we review some basic properties of elliptic curves as well as the elliptic curves specific to the topbox topology, which were previously analysed in~\cite{Adams:2018kez}. For more thorough reviews of elliptic curves and their properties, we refer the reader to one of the various analyses in the literature~\cite{ silverman, Broedel:2017kkb, Adams:2018kez}.  Let us start by considering the generic quartic case of an elliptic curve. An elliptic curve $E$ can be described by the equation
\begin{align}\label{eq:quarticE}
 E \;: \;
 w^2 - \left(z-z_1\right) \left(z-z_2\right) \left(z-z_3\right) \left(z-z_4\right)
 = 0 ,
\end{align}
where the $z_i \in \mathbb{C}$. The $z_i$ define the properties of the elliptic curve and are generally functions of the kinematic variables $x=(x_1,...,x_n)$:
\begin{align}
 z_j \; = \; z_j\left(x\right),
 \;\;\;\;\;\;
 j \in \{1,2,3,4\} \,.
\end{align}
Alternatively, an elliptic curve can be described in terms of two elliptic periods $\psi_i$. In order to define these elliptic periods, let us introduce the auxiliary variables $Z_i$ as
\begin{align}
 Z_1 \; = \; \left(z_2-z_1\right)\left(z_4-z_3\right),
 \;\;\;\;\;\;
 Z_2 \; = \; \left(z_3-z_2\right)\left(z_4-z_1\right),
 \;\;\;\;\;\;
 Z_3 \; = \; \left(z_3-z_1\right)\left(z_4-z_2\right) \,.
\end{align}
The $Z_i$ satisfy the relation
\begin{align}
 Z_1 + Z_2 \; = \; Z_3
\end{align}
and can be used to define the modulus $k^2$ and the complementary modulus $\bar{k}^2$ of the elliptic curve $E$ as 
\begin{equation}
 k^2 
 \; = \; 
 \frac{Z_1}{Z_3},
 \quad
 \bar{k}^2 
 \; = \;
 1 - k^2 
 \; = \;
 \frac{Z_2}{Z_3} \,.
\end{equation}
Then we can choose the two elliptic periods $\psi_i$ associated to the elliptic curve $E$ as
\begin{equation}
\label{def_generic_periods}
 \psi_1 
 \; = \; 
 \frac{4 K\left(k\right)}{Z_3^{\frac{1}{2}}},
 \quad
 \psi_2
 \; = \; 
 \frac{4 i K\left(\bar{k}\right)}{Z_3^{\frac{1}{2}}} \,,
\end{equation}
where $K$ denotes the complete elliptic integral of the first kind
\begin{align}
K(\lambda) = \int_0^1 \frac{d t}{\sqrt{(1-t^2)(1-\lambda t^2)}} \,.
\end{align}

\begin{figure}[t]
  \centering
  \subcaptionbox{The sunrise topology\label{fig:sunrise}}[0.3\textwidth]{\includegraphics[width=2.5cm]{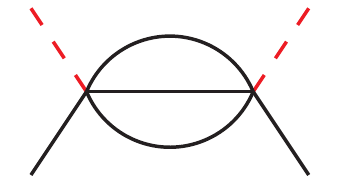}}
  \subcaptionbox{The topbox topology\label{fig:topbox}}[0.3\textwidth]{\includegraphics[width=3cm]{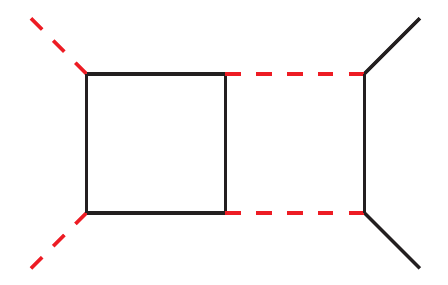}}
  \subcaptionbox{The bubblebox topology\label{fig:bubblebox}}[0.3\textwidth]{\includegraphics[width=2cm]{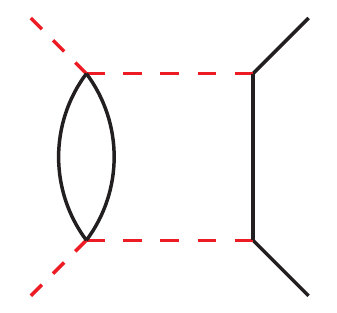}}
\caption{Topologies in the topbox family that are associated with the elliptic curves $E^{(a)}$, $E^{(b)}$ and $E^{(c)}$.
Black-solid lines represent massive particles, red-dashed lines represent massless particles.} \label{fig:subsectors}
\end{figure}

In our results, the dependence of iterated integrals on elliptic curves enters through the
appearance of elliptic periods in the integration kernels Eq.~\eqref{eq:pullback}. Let us now take a
closer look at the different elliptic topologies and the corresponding elliptic curves relevant for
this publication. The topbox diagram has three elliptic sub-sectors corresponding to three different
elliptic curves. They can be obtained from the maximal cuts in the Baikov representation
\cite{Frellesvig_2017,Primo:2016ebd,Primo:2017ipr}. We identify the three different elliptic curves by the labels $a,b,$ and $c$ according to the diagrams depicted in Fig.~\ref{fig:subsectors}. 

The elliptic curve $E^{(a)}$ is associated to the sunrise topology (Fig.~\ref{fig:sunrise}) and is probably the most well-known of the three. The corresponding $z_i$ in Eq.~\eqref{eq:quarticE} are given as
\begin{equation}
 z^{\curveone}_1 \; = \; \frac{t-4m^2}{\mu^2},
 \;\;\;
 z^{\curveone}_2 \; = \; \frac{-m^2-2m\sqrt{t}}{\mu^2},
 \;\;\;
 z^{\curveone}_3 \; = \; \frac{-m^2+2m\sqrt{t}}{\mu^2},
 \;\;\;
 z^{\curveone}_4 \; = \; \frac{t}{\mu^2}.
\end{equation}
Iterated integrals involving only this topology can be cast in terms of iterated integrals of modular forms, and are well-suited for numerical evaluation~\cite{adams2018feynman, brown2017multiple, Broedel:2019kmn, Abreu:2019fgk, Broedel:2017siw, Broedel:2018iwv, Duhr:2019rrs}.

The elliptic curve $E^{(b)}$ is associated to the topbox sector itself (Fig.~\ref{fig:topbox}) and the corresponding $z_i$ are given by
\begin{align}
 z^{\curvetwo}_1 \; & = \; \frac{t-4m^2}{\mu^2},
 \;\;\;
 z^{\curvetwo}_2 \; = \; \frac{-m^2-2m\sqrt{t + \frac{\left(m^2-t\right)^2}{s}}}{\mu^2},
 \;\;\;
 z^{\curvetwo}_3 \; = \; \frac{-m^2+2m\sqrt{t + \frac{\left(m^2-t\right)^2}{s}}}{\mu^2},
 \nonumber \\
 z^{\curvetwo}_4 \; &= \; \frac{t}{\mu^2}.
 \label{curvetwo}
\end{align}

The elliptic curve $E^{(c)}$ is associated to the remaining elliptic sub-sector which we dub the bubblebox sector (Fig.~\ref{fig:bubblebox}). The corresponding quartic is defined by
\begin{align*}
 z^{\curvethree}_1 \; & = \; \frac{t-4m^2}{\mu^2},
 \qquad
 z^{\curvethree}_2 \;  = \; \frac{1}{\mu^2} \left( - m^2 \frac{\left(s+4t\right)}{\left(s-4m^2\right)} - \frac{2}{4m^2-s} \sqrt{s m^2 \left( st + \left(m^2-t\right)^2 \right) } \right),
 \nonumber \\
 z^{\curvethree}_3 \; & = \; \frac{1}{\mu^2} \left( - m^2 \frac{\left(s+4t\right)}{\left(s-4m^2\right)} + \frac{2}{4m^2-s} \sqrt{s m^2 \left( st + \left(m^2-t\right)^2 \right) } \right),
 \qquad
 z^{\curvethree}_4 \;  = \; \frac{t}{\mu^2}.
\end{align*}
A more thorough analysis of the different sub-sectors and the corresponding elliptic curves can be found in \cite{Adams:2018kez}.

\subsection{Results for the new master integrals}

Now that we have identified the various elliptic curves entering the computation, let us turn to the computation of the master integrals $\cJ_{21}$ and $\cJ_{22}$, which are not known from the literature. As we have previously explained, the missing master integrals can be computed order by order in $\epsilon$ from the differential equation given in Eq.~\eqref{eq:DEQ}. We compute the master integrals iteratively in the expansion parameter $\epsilon$ by integrating the differential equation from a base point $(x_0,y_0)= (0,1)$ (corresponding to $s\; =\; \infty$ and $t\;= \; m^2$). The integrals that need to be evaluated receive contributions from the elliptic topbox sub-sectors and hence contain elliptic iterated integrals themselves. The boundary values of the integrals at different orders in $\epsilon$ are inferred from the regular point $x,y=(1,1)$, where these two master integrals vanish. 

The integral $\cJ_{21}$ is associated with elliptic curve $a$, and the first five orders in $\eps$ for $\cJ_{21}$ read
\begin{align}
\cJ_{21}^{(0)} & = 0 \,, \nn 
\cJ_{21}^{(1)} & = 0 \,, \nn 
\cJ_{21}^{(2)} & = 0 \,, \nn 
\cJ_{21}^{(3)} & = 0 \,, \nn 
\cJ_{21}^{(4)} & = -\frac{\pi^4}{60} + \big(G(0,y)-2 G(1,y)\big) \zeta_3 + G(0,0,0,1,y) - 2 G(1,0,0,1,y) \nn \label{eq:J21} 
               & \quad \; + \frac{\pi^2}{36} I_\gamma(g_0,f_3;\lambda) + \frac{1}{18} I_\gamma(g_0,f_3,\eta_0^{(a)},f_3;\lambda) \,.
\end{align}
We have adopted the notation for the integration kernels appearing in the topbox family which was introduced in~\cite{Adams:2018kez}. The integration kernels in Eq.~\eqref{eq:J21} are given by 
\begin{align}
g_0 &= dy \frac{y+3}{y(1-y)} \,, \\
f_3 &= \; 3 \frac{\psi_1^{(a)}}{\pi} dy  \,,\\
\eta_0^{(a)} &= -\frac{2 \pi^2 dy }{{(\psi_1^{(a)})}^2 (-9+y)(-1+y)y  }\,. 
\end{align}

We find that $\cJ_{22}$ contains the elliptic sector 93 from~\cite{Adams:2018kez} in one of its sub-topologies and is hence associated with the elliptic curve $b$. The first four orders in $\eps$ for $\cJ_{21}$ can be written entirely in terms of MPLs and read
\begin{align}
\cJ_{22}^{(0)} & = 0 \,, \nn 
\cJ_{22}^{(1)} & = 0 \,, \nn 
\cJ_{22}^{(2)} & = -\frac{1}{2} G(0,0,x) \,, \nn 
\cJ_{22}^{(3)} & = G(1,y) G(0,0,x) + 3 G(0,-1,0,x) - \frac{3}{2} G(0,0,0,x) + \frac{\pi^2}{4} G(0,x) + \frac{9}{2} \zeta_3 \,.
\end{align}
The explicit dependence on the elliptic curve associated with the topbox sector enters for the first time at $\mathcal{O}(\epsilon^4)$. The corresponding term involves 28 new integration kernels that have not appeared in the computations of the other master integrals and, accordingly, yields a rather large expression. We therefore refrain from showing it here explicitly and refer the interested reader to the ancillary files.

Note that the differential $1$-forms appearing in the iterated integrals are generally not path independent \cite{Brown:2011}. Nevertheless it is clear from a physical point of view that the master integrals do not depend on the particular choice of path chosen for the numerical evaluation of these iterated integrals. We have verified this path independence numerically using the methods explained in the following section.

\section{Numerical evaluations and iterated integrals}
\label{sec:numiterint}
In the previous sections, we discussed the analytic expressions of all the master integrals related
to our process. In this section, we discuss how to numerically compute these integrals. The
integrals containing only MPLs can be evaluated to a high precision, for example, using
\textsc{Ginac}~\cite{Vollinga_2005}. The numerical evaluation of iterated integrals associated with
elliptic curves on the other hand is quite challenging. We explain how to compute these iterated
integrals in two phase space regions, specifically, in the Euclidean and the physical region.
Computing the iterated integrals in the Euclidean region in the small neighbourhood of the boundary
point is relatively straightforward, as was also explained in ~\cite{Adams:2018kez}. For numerically
evaluating these integrals in the physical region we need to analytically continue the associated
elliptic curves across all the branch cuts appropriately. The analytic continuation of integrals
having one parameter dependence has already been discussed in the literature \cite{BOGNER2017528, Abreu:2019fgk, Duhr:2019rrs}.
However, analytically continuing all the integrals in our case, which includes the topbox integral
associated to three different elliptic curves, is much more involved. With the objective of not
digressing too much from the main goal of this paper, we discuss here only the ingredients essential
for this computation.
 
\begin{figure}[H]
    \begin{subfigure}{.5\textwidth}
        \includegraphics[width=\linewidth, right]{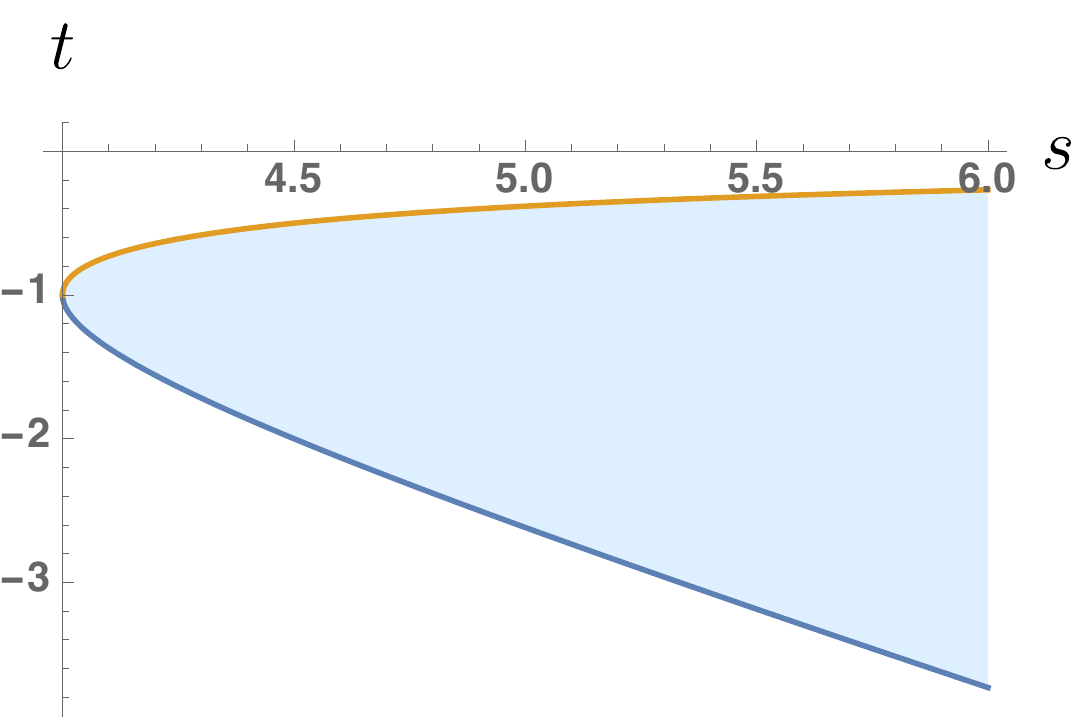}
        \caption{Physical region in $s$ $t$ coordinates}
  \label{fig:sub1}
    \end{subfigure}
    \begin{subfigure}{.5\textwidth}
            \includegraphics[width=\linewidth, left]{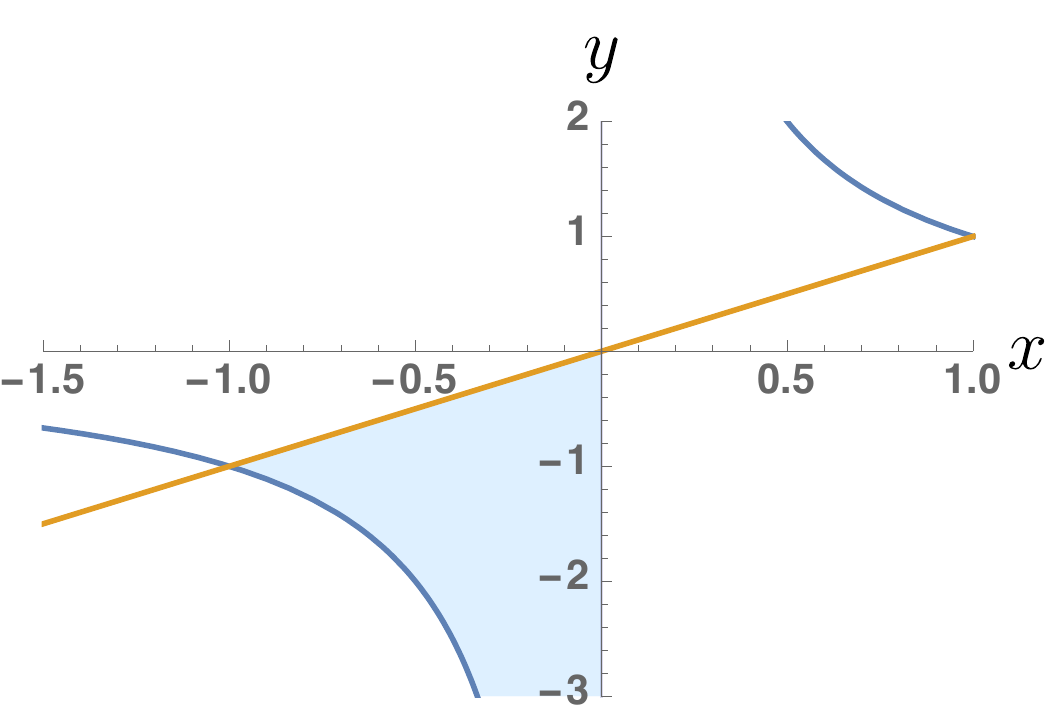}
            \caption{Physical region in $x$ $y$ coordinates}
  \label{fig:sub2}
    \end{subfigure}
    \caption{The physical region for $gg \rightarrow t \bar{t}$.}
    \label{fig:physical}
\end{figure}

The physical region for our case is governed by the following equations ($m_t=1$):
\begin{align*}
 s \geq 4, \qquad G(p_1,p_2,p_3) \geq 0 \; \implies \; s (-t^2 -st+2t-1) \geq 0,
\end{align*}
where $G$ is the Gram-determinant.
In the coordinates $x$ and $y$, these equations take the form
\begin{align*}
& y=t,\\
& x= \frac{1}{2}(2- \sqrt{4-s}\sqrt{-s}-s).
\end{align*}
Fig. \ref{fig:physical} shows the physical region in $s,t$ coordinates as well as $x,y$ coordinates.  

We integrate our iterated integrals from  the base point $(x=0, y=1)$ to any $x$ and $y$ over many
segments and use path decomposition formula to evaluate the contribution over the whole path. The
path segments need to be chosen while taking into consideration both spurious as well as physical
types of singularities in the system. 

The Euclidean region is defined as the region where the iterated integrals have only real
contributions. The numerical evaluation of the iterated integrals that appear in our results for the
master integrals is done for the Euclidean region as follows. We split the integration path from the
boundary point $(0,1)$ to a generic kinematic point $(x,y)$ in the Euclidean region into two pieces and
use the path decomposition formula Eq.~\eqref{eq:path_decomp} to evaluate the integrals.  It is better
to use the coordinate system in Eq.~\ref{eq:xytilde} $(\tilde{x}, \tilde{y})$ here. $\tilde{x}$
rationalizes all the square roots associated with the non-elliptic kernels and $\tilde{y}$ is used
to bring the boundary point of integration to $(0,0)$.  We choose the following paths: $\alpha$, from
$(0,0)$ to $(\tilde{x},0)$ and $\beta$, from $(\tilde{x},0)$ to $(\tilde{x},\tilde{y})$. For all the
master integrals in our case, the integration along the first part gives only MPLs, which can be
evaluated to high precision, as mentioned before. For the integration along the second part, we
expand all integration kernels around $y=1$, assuming $\tilde{y}$ to be small. For integrating to a
point $y$ not close to $1$, we may use more segments along $y$ and use the path decomposition formula recursively.

To evaluate the master integrals in the physical region, we need to analytically continue the
iterated integrals around the physical branch points. We use multiple (one-dimensional) path
segments and use series expansion of the integrands on each of these paths to perform the
integration. The choice of path is controlled by the radius of convergence of the series, which in
turn depends on the singularities present in the kernels. It is important to choose properly the
number of path segments and their sizes, so that the series solution converges properly on each of
these segments and also the computation does not become too slow due to the presence of too many
segments. Generally, the segments should not be larger than half the radius of convergence of the
series expansion \cite{Abreu_2020}. We again use the path decomposition formula to patch together
the contribution from the different segments and compute the iterated integral over the whole path. 

\begin{figure}[H]
  \begin{center}
  \includegraphics[width=0.50\textwidth]{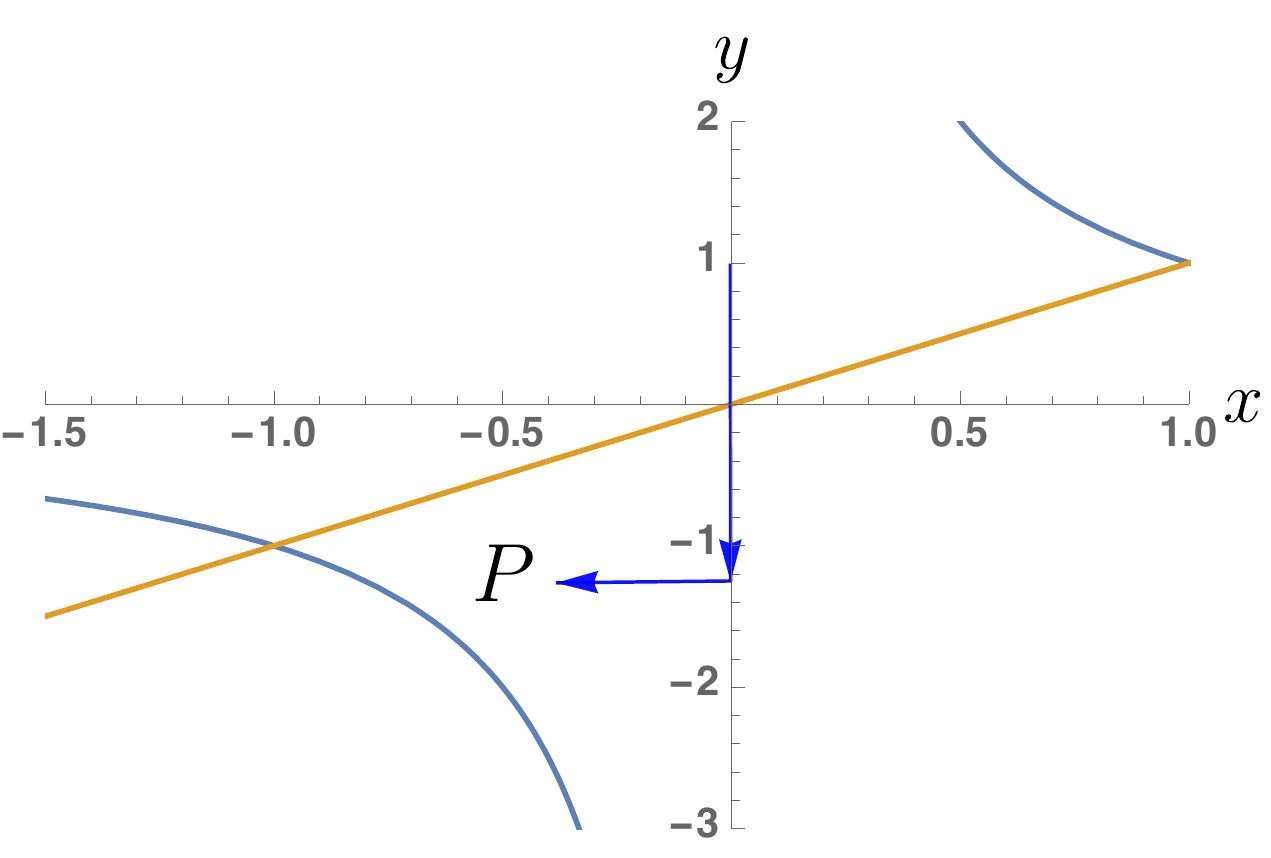}
  \end{center}
  \caption{Integrating to a point $P$ in the physical region.}
  \label{fig:path1}
\end{figure}

For all the sunrise type kernels, there is a singularity while crossing the line $y=0$
($\tilde{y}=1$). For the elliptic kernels having both $x$ and $y$ dependence, there is a singularity
on the line $x=y$, i.e., the line crossing over to the physical region. For the path shown in Fig.
\ref{fig:path1}, both these type of singularities coincide. We therefore find it easier to integrate
all the iterated integrals first along $y$ and then along $x$. The kernels are no longer only MPLs
on either of these paths but evaluate to iterated integrals over elliptic curves.

For all the integrals from the topbox family, we compute analytically in the physical region all the
iterated integrals associated with the curves $a$ and $b$, along with all the integrals depending on
MPLs only. The periods of elliptic curve $c$ are discontinuous on the line $x=0$, unlike periods $a$
and $b$. We reserve for a future publication a detailed explanation of the analytic continuation of
the iterated integrals of the topbox family, which depend on multiple elliptic curves, together with
the study of the analytic continuation of the integrals which depend also on the curve $c$. 

Numerical integration to other regions in the phase space, not discussed above, can also be
performed in a similar way. For example, integrating to the point $(-0.5,0.5)$ in Fig.~\ref{fig:path1} does not require us to cross any physical singularities if we integrate all the
iterated integrals from the base point $(0,1)$. This can also be performed in the same way using the
path decomposition formula recursively and does not involve a lot of complications. Another way to
perform the integration in different phase space points is to start integrating the iterated
integrals from a point already in that region. While this avoids the problem of analytic
continuation it requires us to compute the boundary conditions at the new base points.

\subsection{Functional Relations Among Iterated Integrals}

In the previous sections, we have presented how we compute the planar master integrals contributing to the $gg \rightarrow t \bar{t}$ scattering process in terms of MPLs and iterated integrals over elliptic curves. While this allows us to evaluate the $t\bar{t}gg$ helicity amplitudes, we find that there are potential redundancies in the functional basis of iterated integrals that might make the expression appear complex. Indeed, after integrating the differential equation, we found that the simple pole vanished numerically as expected but this cancellation was not reflected in the analytic expression. In order to achieve an explicit cancellation of all pole terms, some functional relations among the iterated integrals had to be applied. In this section we comment on functional relations among iterated integrals and the cancellation of the $\epsilon$ pole. For this purpose, we adopt again the notation for the integration kernels introduced in~\cite{Adams:2018kez}. \\\

The functional relations we will discuss in this section are related to the appearance of derivatives of elliptic periods in the integration kernels. Take for example the kernel 
\begin{align} 
a^{(b)}_{3,3} &= \frac{ \psi_1^{(b)} x}{\pi  (x-1)^2} dy -\frac{(\partial_y \psi_1^{(b)})  x (y-1)^2 \left(x^2 y+3 x^2-6 x y-2 x+y+3\right)}{\pi  (x-1)^3 (x+1) \left(3 x^2-2 x y-4 x+3\right)} dx \\ \nonumber
&+ \frac{(\partial_y \psi_1^{(b)}) x (y-1)}{\pi  (x-1)^2} dy-\frac{ \psi_1^{(b)} (y-1) \left(3 x^4-x^3 y+x^3-6 x^2 y-x y+x+3\right)}{\pi  (x-1)^3 (x+1) \left(3 x^2-2 x y-4 x+3\right)} dx \\
&= \left(\frac{(\partial_x \psi_1^{(b)}) x (y-1)}{\pi  (x-1)^2}-\frac{\psi_1^{(b)} (x+1) (y-1)}{\pi  (x-1)^3}\right) dx \\ \nonumber
&+\left(\frac{(\partial_y \psi_1^{(b)}) x (y-1)}{\pi  (x-1)^2}+\frac{\psi_1^{(b)} x}{\pi  (x-1)^2}\right) dy\\
&= d \left( \psi_1^{(b)} \frac{ x (-1 + y)}{\pi (-1 + x)^2}\right) \,.
\end{align} 

We see that the integration kernel $a^{(b)}_{3,3} $ corresponds to a total derivative. Because the primitive of $a^{(b)}_{3,3} $ vanishes at the lower integration boundary, we have

\begin{align}
I(a^{(b)}_{3,3}) = \psi_1^{(b)} \frac{ x (-1 + y)}{\pi (-1 + x)^2} \,.
\end{align}

We can furthermore use integration by parts to derive other relations for iterated integrals involving the kernel $a^{(b)}_{3,3}$. In particular, we have
\begin{align}
I(a^{(b)}_{3,3},f,\dots ) &= \int a^{(b)}_{3,3} I(f,\dots ) \\ \label{eq:totDiffIntegral}
& = \int d\left( \psi_1^{(b)} \frac{ x (-1 + y)}{\pi (-1 + x)^2}\right) \cdot I(f,\dots ) \\ \label{eq:iterintRelations}
& = \left[  \psi_1^{(b)} \frac{ x (-1 + y)}{\pi (-1 + x)^2} I(f,\dots) \right]_{(0,1)}^{(x,y)} - I\left(  \psi_1^{(b)} \frac{ x (-1 + y)}{\pi (-1 + x)^2} f,\dots\right) \,,
\end{align}
where $f$ is an arbitrary integration kernel. Note that the first factor in the integral in Eq.~\eqref{eq:totDiffIntegral} denotes a total differential and not the integration measure. If we can recast the product 
\begin{align}
 \psi_1^{(b)} \frac{ x (-1 + y)}{\pi (-1 + x)^2} f
\end{align}
as a linear combination of integration kernels, we have successfully identified a relation among iterated integrals $I$ which is closed under the minimal set of integration kernels we are using. We find the following equalities
\begin{align}
\psi_1^{(b)} \frac{ x (-1 + y)}{\pi (-1 + x)^2}  \eta^{(b)}_{0} &= \frac{2}{3} \eta^{(b)}_{1,4} \,, \\
\psi_1^{(b)} \frac{ x (-1 + y)}{\pi (-1 + x)^2}  \eta^{(a,b)}_{2} &= \frac{1}{72} \eta^{(a)}_{3,2} - \frac{4}{27} f_3 \,, \\
\psi_1^{(b)} \frac{ x (-1 + y)}{\pi (-1 + x)^2} \eta^{(b)}_{1,1} &= \frac{3}{2} \eta^{(b)}_{2,1} - \frac{1}{8}\omega_0 + \frac{1}{8}\omega_4 \,.
\end{align}
Plugging the above linear combinations into Eq.~\eqref{eq:iterintRelations}, we obtain three functional relations which we can use to simplify the pole terms. After doing so we are left with iterated integrals involving only rational kernels. These integrals evaluate to MPLs and cancel the ones already present in the expression, leading to the exact cancellation of the pole terms. 

We can also use these relations to reduce the number of elements in the monomial basis of special
function in the two-loop finite remainders with a closed top-quark loop $A^{(2),N_h}$.
When mapping the master integrals appearing in $A^{(2),N_h}$ to a basis of special functions, we found that there were 12025 monomials which, applying
relations involving $a^{(b)}_{3,3}$ as described above, was reduced to 11791 monomials.
We notice that there are much fewer monomial basis elements appearing in the two-loop finite remainders $F^{(2),N_h}$.
For example, for one of the subamplitudes with helicities $+++-$, only 3586 monomial basis elements appear in the finite remainder with non-zero coefficients.
Applying the relations discussed above we find that this number is reduced to 3158.
Similar situations are also observed for the other subamplitudes and the $++++$ helicity configuration.

Note also that there are more kernels containing derivatives of elliptic periods. 
It is possible to write some of these kernels as a sum of a total derivative which captures all appearing derivatives of elliptic periods 
and a new kernel which does not contain the derivative of an elliptic period. 
By doing so, it might be possible to find further relations. 
Because of the large number of iterated integrals contributing to the scattering amplitude and the need to introduce new integration kernels, 
from our experience with $a^{(b)}_{3,3}$ kernel, finding more relations between iterated integrals is necessary to simplify the finite remainder.
We therefore reserve a thorough analysis of the possible functional relations for another publication.

\section{Numerical Results}
For benchmarking purposes, we present numerical results at a phase space point.
We use a Euclidean phase space point that is considered in~\cite{Adams:2018kez}
\begin{align}
p_{1}&=(-144.17208, 1.5442404, 0.34424036 i, 144.13300),\nonumber \\
p_{2}&=(75.448430, 112.32514, -113.00985 i, -76.405203),\nonumber \\
p_{3}&=(115.65233, -114.70819, 113.70819 i, -114.66056),\nonumber \\
p_{4}&=(-46.928678, 0.83881549, -1.0425849 i, 46.932762),\nonumber \\
m_t &= 3, 
\label{eq:topboxPSpoint}
\end{align}
with the following choice of reference vectors 
\begin{align}
\eta_{1}&=(74.948430, 112.00816, -113.32683 i, -76.905203),\nonumber \\
\eta_{2}&=(-29.015635, 1.5442404, 0.34424036 i, 28.976558).\nonumber
\end{align}
The rational representation of the momenta listed above is provided in the ancillary files.
The set of momenta given in Eq.~\eqref{eq:topboxPSpoint} corresponds to the following Mandelstam invariants
\begin{equation}
s = -\frac{38307}{280}, \qquad t = \frac{90}{11},  \nonumber
\end{equation}
and
\begin{equation}
x = \frac{7}{120}, \qquad y = \frac{10}{11}, \qquad \tilde{x}=\frac{1}{15},\qquad \tilde{y}=\frac{1}{11}.  \nonumber
\end{equation}

\begin{table}[t]
\begin{center}
\small
\begin{tabularx}{0.95\textwidth}{|C{0.7}|C{0.5}|C{1.1}|C{1.2}|C{1.2}|C{1.3}|}
  \hline
   & $\eps^{-4}$ & $\eps^{-3}$ & $\eps^{-2}$ & $\eps^{-1}$ & $\eps^{0}$ \\
  \hline
  $\widehat{A}^{(2),1}_{++++}$ & 2 & 1.315500267 & 174.3336185 & 804.4569447 & -838.7774702 \\
  $\widehat{A}^{(2),1}_{+++-}$ & 2 & 1.315500267 & 16.27135874 & 69.36710669 & -252.9757059 \\
  \hline
  $\widehat{A}^{(2),N_l}_{++++}$ & 0 & 0.3333333333 & 1.200099178 & 60.28730935 & 752.8844277 \\
  $\widehat{A}^{(2),N_l}_{+++-}$ & 0 & 0.3333333333 & 1.080361156 & 7.611962871 & 69.15926288 \\
  \hline
  $\widehat{A}^{(2),N_h}_{++++}$ & 0 & 1.333333333 & 1.591367069 & 59.7088878 & 334.5603307 \\
  $\widehat{A}^{(2),N_h}_{+++-}$ & 0 & 1.333333333 & 1.5496112 & 7.095183901 &  34.06132973 \\
  \hline
  $\widehat{A}^{(2),N_l^2}_{++++}$ & 0 & 0 & 0 & 0.03991267399 & -0.004367011612 \\
  $\widehat{A}^{(2),N_l^2}_{+++-}$ & 0 & 0 & 0 & 0 & 0 \\
  \hline
  $\widehat{A}^{(2),N_l N_h}_{++++}$ & 0 & 0 & 0 & 0.05383129703 & -0.04163962585 \\
  $\widehat{A}^{(2),N_l N_h}_{+++-}$ & 0 & 0 & 0 & 0 & 0 \\
  \hline
  $\widehat{A}^{(2),N_h^2}_{++++}$ & 0 & 0 & 0 & 0.01391862304 & -0.02907884376 \\
  $\widehat{A}^{(2),N_h^2}_{+++-}$ & 0 & 0 & 0 & 0 & 0 \\
  \hline
\end{tabularx}
\end{center}
\caption{Mass renormalised two-loop amplitudes (normalised to the tree level amplitude) using the kinematic point in Eq.~\eqref{eq:topboxPSpoint}.
}
\label{tab:num2LmrenPSeuclidean}
\end{table}
We present results for the mass renormalised two-loop helicity amplitude for various fermion loop contributions specified in Eq.~\eqref{eq:flamplitude2loop},
normalised to the tree-level amplitude
\begin{equation}
\hat{A}^{(2),f}_{\lambda_1 \lambda_2 \lambda_3 \lambda_4} = 
\frac{A^{(2),f}(1_{\bar t}^{\lambda_1},2_{t}^{\lambda_2},3_{g}^{\lambda_3},4_{g}^{\lambda_4} )}{
A^{(0)}(1_{\bar t}^{\lambda_1},2_{t}^{\lambda_2},3_{g}^{\lambda_3},4_{g}^{\lambda_4} )
},
\end{equation}
with helicities $\lambda_i$ and $f = 1, N_l, N_h, N_l^2, N_l N_h, N_h^2$.
Numerical results at the kinematic point given in Eq.~\eqref{eq:topboxPSpoint} are displayed in Table~\ref{tab:num2LmrenPSeuclidean}.
We use \textsc{Ginac} to evaluate MPLs that appear in the amplitude, while the numerical evaluation of iterated integrals involving elliptic
kernels are done using the method described in Section~\ref{sec:numiterint}.

We have performed several checks to validate the analytic results derived in this paper:
\begin{itemize}

\item We compare numerical results in the Euclidean region for the squared matrix element obtained from the helicity amplitudes against the squared matrix element
derived from an independent computation that directly computes the interference between the tree level and the two loop amplitudes in the HV scheme.

\item We compare the finite remainder of the squared matrix element obtained from the helicity amplitudes against the results of~\cite{Baernreuther:2013caa}
at the following physical point
\begin{equation}
\frac{s}{m_t^2} = 5, \qquad \frac{t}{m_t^2} = -\frac{5}{4}, \qquad \mu = m_t, \qquad m_t=1,
\end{equation}
with $N_l+N_h$ active flavours in the running of $\alpha_s$ has been used.
In order to extract the finite remainder with $N_l+N_h$ active flavours from~\cite{Baernreuther:2013caa}, we subtract the IR poles in the CDR scheme from 
the renormalised amplitude given in Table~3 of~\cite{Baernreuther:2013caa}. 
As explained in Section~\ref{sec:numiterint}, in this paper we have considered evaluation of the iterated integrals associated
with the curves $a$ and $b$  in the physical region from analytical results while leaving the study of the analytic continuation of the integrals
which depend also on the curve $c$ for the future publication. 
Therefore, for the evaluation of the amplitude with a single closed top-loop contribution ($A^{(2),N_h}$) in the physical region, 
we obtain numerical results of the mass renormalised amplitude in the master integral representation (Eq.~\eqref{eq:calc_ampMI}).
In this representation, the master integrals containing elliptic curve $c$ are evaluated using \textsc{Fiesta}~\cite{Smirnov:2015mct} 
and \textsc{pySecDec}~\cite{Borowka:2017idc}, while the remaining master integrals are evaluated from their analytical expressions.
From this master integral representation we further remove the UV and IR poles numerically to obtain the finite remainder.

\end{itemize}

\section{Conclusions}

In this article we have computed a set of analytic helicity amplitudes for the planar corrections to
top quark pair production via gluon fusion at two loops in QCD. While these amplitudes have been
available for some time in alternative forms, this is the first time that amplitude level expressions
using the massive spinor-helicity formalism have been obtained. We demonstrate that this form
is suitable for processing with a rational parametrisation of the kinematics. This has the benefit
that the algebra can be performed using finite field arithmetic and so combat the growth in
algebraic complexity. This method will scale better than conventional approaches when considering
additional final state particles.

Another important new ingredient is the inclusion of top quark loops using analytic expressions for
the master integrals. Firstly we observed that the complete set of master integrals contributing to
the amplitude were not available in the literature and computed them using the differential equation
method. These integrals give rise to a set of special functions which have recently
been presented in terms of iterated integrals over kernels including three elliptic curves. The
growth in analytic complexity encountered in this sector is considerable and we presented a study of
the numerical evaluation as well as analytic simplification leading to complete cancellation of the
universal IR and UV poles. 

We provide a complete set of independent finite remainders in the ancillary files accompanying the
arXiv version of this article. We have also performed a cross check on the ingredients in our
computation against the available numerical results.

Our work leaves a few open questions that motivate further investigation. The numerical evaluation
of the iterated integrals is not at the same level of maturity as the commonly used MPLs. In
particular the analytic continuation allowing a stable and efficient evaluation over the whole
physical scattering region requires further study. We also notice that the iterated integrals obey
additional functional relations. A complete understanding of these identities could and lead to a
substantial reduction in the complexity of the final expressions and deserves additional study.

\begin{acknowledgments}
We thank Matteo Becchetti and Stefan Weinzierl for helpful discussions.
This project has received funding from the European Union’s Horizon 2020 research and innovation programmes 
\textit{New level of theoretical precision for LHC Run 2 and beyond} (grant agreement No 683211) and
\textit{High precision multi-jet dynamics at the LHC} (grant agreement No 772009).
HBH has been partially supported by STFC consolidated HEP theory grant ST/T000694/1.
\end{acknowledgments}

\appendix

\section{Renormalisation constants}
\label{app:renormalisation}

In this Appendix, we provide renormalisation constants required to derive the renormalised amplitudes defined in 
Eqs.~\eqref{eq:ampren_1loop}~and~\eqref{eq:ampren_2loop}. Their expansion in the bare strong coupling constant $\alpha_s^0$ are

\begin{align}
Z_m = \; & 1 + \bigg(\frac{\alpha_s^0}{4\pi}\bigg) \delta Z_m^{(1)} +  \bigg(\frac{\alpha_s^0}{4\pi}\bigg)^2 \delta Z_m^{(2)} + \cO\big((\alpha_s^0)^3\big), \\
Z_t = \; & 1 + \bigg(\frac{\alpha_s^0}{4\pi}\bigg) \delta Z_t^{(1)} +  \bigg(\frac{\alpha_s^0}{4\pi}\bigg)^2 \delta Z_t^{(2)} + \cO\big((\alpha_s^0)^3\big), \\
Z_g = \; & 1 + \bigg(\frac{\alpha_s^0}{4\pi}\bigg) \delta Z_g^{(1)} +  \bigg(\frac{\alpha_s^0}{4\pi}\bigg)^2 \delta Z_g^{(2)} + \cO\big((\alpha_s^0)^3\big), \\
\alpha_s^0 = \; & \alpha_s \; \bigg\lbrace 1 + \bigg(\frac{\alpha_s}{4\pi}\bigg) \delta Z_{\alpha_s}^{(1)} 
   +  \bigg(\frac{\alpha_s}{4\pi}\bigg)^2 \delta Z_{\alpha_s}^{(2)} \bigg\rbrace  + \cO(\alpha_s^4).
\end{align}
We recompute the following $d_s$ dependent mass counterterms in the integral representation that contribute at leading colour 
\begin{align}
\delta Z_m^{(1)} = \; & N_c \; \frac{2+d_s-2d_s \eps }{4 (1-2\eps) m_t^2} \; I\big(\usepix{0.6cm}{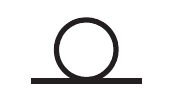}{7}{0}{7}{0}\big), 
   \label{eq:dZm1} \\
\delta Z_m^{(2)} = \; & N_c^2 \; \bigg\lbrace 
    \frac{1}{m_t^4}\bigg( \frac{1-6\eps+6\eps^2}{4(1-2\eps)^2} - d_s \frac{2-3\eps}{4(1-2\eps)} + \frac{d_s^2}{16} \bigg) 
    \; I\big(\usepix{0.6cm}{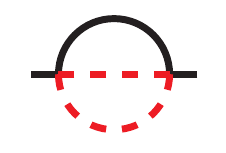}{8}{0}{8}{0}\big) \nn
 & \qquad + \frac{1}{m_t^2 (1-\eps)}\bigg( -\frac{7-16\eps+12\eps^3}{4(1-2\eps)(1-4\eps)} - d_s \frac{1-8\eps+10\eps^2}{4(1-4\eps)} + d_s^2 \frac{1-2\eps}{16} \bigg)
     I\big(\usepix{0.5cm}{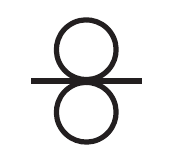}{6}{0}{6}{0}\big) \bigg\rbrace\nn
 & + N_c N_l \; \frac{-d_s-6\eps+4d_s\eps}{4(1-4\eps) m_t^2} \; I\big(\usepix{0.6cm}{figs/sunriseM00}{8}{0}{8}{0}\big) \nn
 & + N_c N_h \; \bigg\lbrace -\frac{3(1-\eps)(-1-4\eps+d_s\eps+\eps^2+d_s\eps^2)}{4\eps(1+\eps)(1-2\eps) m_t^4} 
             \; I\big(\usepix{0.5cm}{figs/tadpole2}{6}{0}{6}{0}\big) \nn
 & \hspace{2cm} + \frac{-2-3\eps+d_s\eps+3\eps^2+d_s\eps^2}{4\eps(1+\eps)m_t^2} \; I\big(\usepix{0.6cm}{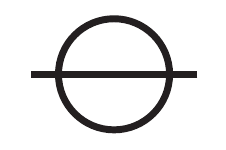}{8}{0}{8}{0}\big) \bigg\rbrace,
  \label{eq:dZm2} 
\end{align}
where the tadpole and on-shell sunrise integrals are defined by
\begin{align}
I\big(\usepix{0.6cm}{figs/tadpole}{7}{0}{7}{0}\big)    = \; & \left(m_t^2\right)^\eps \int \dk{1} \; \frac{1}{k_1^2-m_t^2}, \\
I\big(\usepix{0.5cm}{figs/tadpole2}{6}{0}{6}{0}\big)   = \; & \left(m_t^2\right)^{2\eps} \int \dk{1}\dk{2} \; \frac{1}{(k_1^2-m_t^2) (k_2^2-m_t^2)}, \\
I\big(\usepix{0.6cm}{figs/sunriseM00}{8}{0}{8}{0}\big) = \; & \left(m_t^2\right)^{2\eps} \int \dk{1}\dk{2}  
                                                         \; \frac{1}{(k_1^2-m_t^2) k_2^2 (k_1+k_2+p)^2}, \label{eq:OSsunriseM00}\\
I\big(\usepix{0.6cm}{figs/sunriseMMM}{8}{0}{8}{0}\big) = \; & \left(m_t^2\right)^{2\eps} \int \dk{1}\dk{2}  
                                                         \; \frac{1}{(k_1^2-m_t^2) (k_2^2-m_t^2) ((k_1+k_2+p)^2-m_t^2)} \label{eq:OSsunriseMMM}, 
\end{align}
where $p^2 = m_t^2$ in Eqs.~\eqref{eq:OSsunriseM00}~and~\eqref{eq:OSsunriseMMM}. 
The normalisation of the integrals appearing in Eqs.~\eqref{eq:dZm1}~and~\eqref{eq:dZm2}
is chosen such that it matches the convention adopted in Eq.~\eqref{eq:colourdecomposition}.
We have checked that the integral representation of mass counterterms reproduce the known integrated forms given in Ref.~\cite{Melnikov:2000zc}.
The wavefunction and strong coupling renormalisation constants at one loop are~\cite{Bonciani:2010mn}
\begin{align}
\delta Z_t^{(1)} = \; & \ceps (m_t^2)^{-\eps} \; C_F \; \bigg( -\frac{3}{\eps} - \frac{4}{1-2\eps} \bigg), \\
\delta Z_g^{(1)} = \; & \ceps (m_t^2)^{-\eps} \; T_F N_h \; \bigg( - \frac{4}{3\eps} \bigg), \\
\delta Z_{\as}^{(1)} = \; & \ceps \; \frac{e^{-\gamma_E}}{\Gamma(1+\eps)}\;\bigg(-\frac{\beta_0}{\eps}\bigg),
\end{align}
and two loops~\cite{Bonciani:2010mn,Czakon:2007ej,Czakon:2007wk}
\begin{align}
\delta Z_t^{(2)} = \; & \ceps^2 (m_t^2)^{-2\eps} \; C_F \; \bigg\lbrace
   C_F \bigg(\frac{9}{2\eps^2} + \frac{51}{4\eps} + \frac{433}{8} - 24 \zeta(3) + 96 \zeta(2)\log(2) - 78\zeta(2) \bigg) \nn
& + C_A \bigg( -\frac{11}{2\eps^2}-\frac{101}{4\eps}-\frac{803}{8}+12\zeta(3)-48\zeta(2)\log(2)+30\zeta(2) \bigg) \nn
& + N_l T_F \bigg( \frac{2}{\eps^2} + \frac{9}{\eps} + 8\zeta(2) + \frac{59}{2} \bigg) \nn
& + N_h T_F  \bigg( \frac{4}{\eps^2} + \frac{19}{3\eps} + \frac{1139}{18} - \frac{16\zeta(2)^2}{3} \bigg) + \cO(\eps) \bigg\rbrace, \\
\delta Z_g^{(2)} = \; & \ceps^2 (m_t^2)^{-2\eps} \; T_F N_h \; \bigg\lbrace C_F \; \frac{4(-1-7\eps+4\eps^3)}{\eps(2-\eps-8\eps^2+4\eps^3)} \nn
& + C_A \; \frac{2(-3 - 11\eps + \eps^2 + 15\eps^3 - 4\eps^5)}{\eps^2 (6 + 7\eps - 13\eps^2 - 4\eps^3 + 4\eps^4)} 
  + N_h T_F \; \frac{16}{9\eps^2} \bigg\rbrace, \\
\delta Z_{\as}^{(2)}  = \; &  \ceps^2 \; \frac{e^{-2\gamma_E}}{\Gamma^2(1+\eps)}\;\bigg(\frac{\beta_0^2}{\eps^2} - \frac{\beta_1}{2\eps} \bigg),
\end{align}
where $\ceps = (4\pi)^\eps \Gamma(1+\eps)$. The beta function coefficients are 
\begin{align}
\beta_0 = \; & \frac{11}{3}C_A-\frac{4}{3} T_F (N_l+N_h),\\
\beta_1 = \; & \frac{34}{3}C_A^2 -\frac{20}{3} C_A T_F (N_l+N_h) -4 C_F T_F (N_l+N_h).
\end{align}

\section{Numerical results for one-loop amplitudes}

We present in Table~\ref{tab:num1LmrenPSeuclidean} numerical results for one-loop mass-renormalised amplitude at the kinematic 
point given in Eq.~\eqref{eq:topboxPSpoint}, evaluated through $\cO(\eps^2)$ for various fermion loop contributions defined in Eq.~\eqref{eq:flamplitude1loop}. 
These one-loop amplitudes are required in the computation of the pole terms in Eqs.~\eqref{eq:poleexpandedUV}~and~\eqref{eq:poleexpandedIR}.

\begin{table}[t]
\begin{center}
\small
\begin{tabularx}{0.95\textwidth}{|C{0.7}|C{0.6}|C{1.1}|C{1.2}|C{1.2}|C{1.2}|}
  \hline
   & $\eps^{-2}$ & $\eps^{-1}$ & $\eps^{0}$ & $\eps^{1}$ & $\eps^{2}$ \\
  \hline
  $\widehat{A}^{(1),1}_{++++}$ & -2 & -1.5744168 & -89.02897164 & -497.5712587 & -1803.749249 \\
  $\widehat{A}^{(1),1}_{+++-}$ & -2 & -1.5744168 & -9.997841784 & -47.34990423 & -200.9328154 \\
  \hline
  $\widehat{A}^{(1),N_l}_{++++}$ & 0 & 0 & -0.05986901099 & -0.05659375228 & -0.140485301 \\
  $\widehat{A}^{(1),N_l}_{+++-}$ & 0 & 0 & 0 & 0 & 0 \\
  \hline
  $\widehat{A}^{(1),N_h}_{++++}$ & 0 & 0 & -0.02087793457 & 0.01501707738 & -0.02381992044 \\
  $\widehat{A}^{(1),N_h}_{+++-}$ & 0 & 0 & 0 & 0 & 0 \\
  \hline
\end{tabularx}
\end{center}
\caption{Mass renormalised one-loop amplitudes (normalised to the tree level amplitude) using the kinematic point in Eq.~\eqref{eq:topboxPSpoint}.
}
\label{tab:num1LmrenPSeuclidean}
\end{table}

\bibliographystyle{JHEP}
\bibliography{planar_ttgg.bib}

\providecommand{\href}[2]{#2}\begingroup\raggedright\begin{thebibliography}{100}

\bibitem{Nason:1987xz}
P.~Nason, S.~Dawson and R.~K. Ellis, \emph{{The Total Cross-Section for the
  Production of Heavy Quarks in Hadronic Collisions}},
  \href{http://dx.doi.org/10.1016/0550-3213(88)90422-1}{\emph{Nucl. Phys. B}
  {\bf 303} (1988) 607--633}.

\bibitem{Nason:1989zy}
P.~Nason, S.~Dawson and R.~K. Ellis, \emph{{The One Particle Inclusive
  Differential Cross-Section for Heavy Quark Production in Hadronic
  Collisions}},
  \href{http://dx.doi.org/10.1016/0550-3213(89)90286-1}{\emph{Nucl. Phys. B}
  {\bf 327} (1989) 49--92}.

\bibitem{Baernreuther:2012ws}
P.~B\"arnreuther, M.~Czakon and A.~Mitov, \emph{{Percent Level Precision
  Physics at the Tevatron: First Genuine NNLO QCD Corrections to $q \bar{q} \to
  t \bar{t} + X$}},
  \href{http://dx.doi.org/10.1103/PhysRevLett.109.132001}{\emph{Phys. Rev.
  Lett.} {\bf 109} (2012) 132001}, [\href{http://arxiv.org/abs/1204.5201}{{\tt
  1204.5201}}].

\bibitem{Czakon:2012zr}
M.~Czakon and A.~Mitov, \emph{{NNLO corrections to top-pair production at
  hadron colliders: the all-fermionic scattering channels}},
  \href{http://dx.doi.org/10.1007/JHEP12(2012)054}{\emph{JHEP} {\bf 12} (2012)
  054}, [\href{http://arxiv.org/abs/1207.0236}{{\tt 1207.0236}}].

\bibitem{Czakon:2012pz}
M.~Czakon and A.~Mitov, \emph{{NNLO corrections to top pair production at
  hadron colliders: the quark-gluon reaction}},
  \href{http://dx.doi.org/10.1007/JHEP01(2013)080}{\emph{JHEP} {\bf 01} (2013)
  080}, [\href{http://arxiv.org/abs/1210.6832}{{\tt 1210.6832}}].

\bibitem{Czakon:2013goa}
M.~Czakon, P.~Fiedler and A.~Mitov, \emph{{Total Top-Quark Pair-Production
  Cross Section at Hadron Colliders Through $O(\alpha^4_S)$}},
  \href{http://dx.doi.org/10.1103/PhysRevLett.110.252004}{\emph{Phys. Rev.
  Lett.} {\bf 110} (2013) 252004}, [\href{http://arxiv.org/abs/1303.6254}{{\tt
  1303.6254}}].

\bibitem{Czakon:2015owf}
M.~Czakon, D.~Heymes and A.~Mitov, \emph{{High-precision differential
  predictions for top-quark pairs at the LHC}},
  \href{http://dx.doi.org/10.1103/PhysRevLett.116.082003}{\emph{Phys. Rev.
  Lett.} {\bf 116} (2016) 082003}, [\href{http://arxiv.org/abs/1511.00549}{{\tt
  1511.00549}}].

\bibitem{Czakon:2010td}
M.~Czakon, \emph{{A novel subtraction scheme for double-real radiation at
  NNLO}}, \href{http://dx.doi.org/10.1016/j.physletb.2010.08.036}{\emph{Phys.
  Lett. B} {\bf 693} (2010) 259--268},
  [\href{http://arxiv.org/abs/1005.0274}{{\tt 1005.0274}}].

\bibitem{Bonciani:2015sha}
R.~Bonciani, S.~Catani, M.~Grazzini, H.~Sargsyan and A.~Torre, \emph{{The $q_T$
  subtraction method for top quark production at hadron colliders}},
  \href{http://dx.doi.org/10.1140/epjc/s10052-015-3793-y}{\emph{Eur. Phys. J.
  C} {\bf 75} (2015) 581}, [\href{http://arxiv.org/abs/1508.03585}{{\tt
  1508.03585}}].

\bibitem{Catani:2019iny}
S.~Catani, S.~Devoto, M.~Grazzini, S.~Kallweit, J.~Mazzitelli and H.~Sargsyan,
  \emph{{Top-quark pair hadroproduction at next-to-next-to-leading order in
  QCD}}, \href{http://dx.doi.org/10.1103/PhysRevD.99.051501}{\emph{Phys. Rev.
  D} {\bf 99} (2019) 051501}, [\href{http://arxiv.org/abs/1901.04005}{{\tt
  1901.04005}}].

\bibitem{Catani:2019hip}
S.~Catani, S.~Devoto, M.~Grazzini, S.~Kallweit and J.~Mazzitelli,
  \emph{{Top-quark pair production at the LHC: Fully differential QCD
  predictions at NNLO}},
  \href{http://dx.doi.org/10.1007/JHEP07(2019)100}{\emph{JHEP} {\bf 07} (2019)
  100}, [\href{http://arxiv.org/abs/1906.06535}{{\tt 1906.06535}}].

\bibitem{Catani:2020tko}
S.~Catani, S.~Devoto, M.~Grazzini, S.~Kallweit and J.~Mazzitelli,
  \emph{{Top-quark pair hadroproduction at NNLO: differential predictions with
  the $\overline{MS}$ mass}},
  \href{http://dx.doi.org/10.1007/JHEP08(2020)027}{\emph{JHEP} {\bf 08} (2020)
  027}, [\href{http://arxiv.org/abs/2005.00557}{{\tt 2005.00557}}].

\bibitem{Czakon:2008zk}
M.~Czakon, \emph{{Tops from Light Quarks: Full Mass Dependence at Two-Loops in
  QCD}}, \href{http://dx.doi.org/10.1016/j.physletb.2008.05.028}{\emph{Phys.
  Lett. B} {\bf 664} (2008) 307--314},
  [\href{http://arxiv.org/abs/0803.1400}{{\tt 0803.1400}}].

\bibitem{Baernreuther:2013caa}
P.~Bärnreuther, M.~Czakon and P.~Fiedler, \emph{{Virtual amplitudes and
  threshold behaviour of hadronic top-quark pair-production cross sections}},
  \href{http://dx.doi.org/10.1007/JHEP02(2014)078}{\emph{JHEP} {\bf 02} (2014)
  078}, [\href{http://arxiv.org/abs/1312.6279}{{\tt 1312.6279}}].

\bibitem{Chen:2017jvi}
L.~Chen, M.~Czakon and R.~Poncelet, \emph{{Polarized double-virtual amplitudes
  for heavy-quark pair production}},
  \href{http://dx.doi.org/10.1007/JHEP03(2018)085}{\emph{JHEP} {\bf 03} (2018)
  085}, [\href{http://arxiv.org/abs/1712.08075}{{\tt 1712.08075}}].

\bibitem{Adams_2017}
L.~Adams, E.~Chaubey and S.~Weinzierl, \emph{Simplifying differential equations
  for multiscale feynman integrals beyond multiple polylogarithms},
  \href{http://dx.doi.org/10.1103/physrevlett.118.141602}{\emph{Physical Review
  Letters} {\bf 118} (Apr, 2017) }.

\bibitem{Adams:2018bsn}
L.~Adams, E.~Chaubey and S.~Weinzierl, \emph{{Planar Double Box Integral for
  Top Pair Production with a Closed Top Loop to all orders in the Dimensional
  Regularization Parameter}},
  \href{http://dx.doi.org/10.1103/PhysRevLett.121.142001}{\emph{Phys. Rev.
  Lett.} {\bf 121} (2018) 142001}, [\href{http://arxiv.org/abs/1804.11144}{{\tt
  1804.11144}}].

\bibitem{Adams:2018kez}
L.~Adams, E.~Chaubey and S.~Weinzierl, \emph{{Analytic results for the planar
  double box integral relevant to top-pair production with a closed top loop}},
  \href{http://dx.doi.org/10.1007/JHEP10(2018)206}{\emph{JHEP} {\bf 10} (2018)
  206}, [\href{http://arxiv.org/abs/1806.04981}{{\tt 1806.04981}}].

\bibitem{adams2018feynman}
L.~Adams and S.~Weinzierl, \emph{Feynman integrals and iterated integrals of
  modular forms},  2018.

\bibitem{Abreu_2020}
S.~Abreu, H.~Ita, F.~Moriello, B.~Page, W.~Tschernow and M.~Zeng,
  \emph{Two-loop integrals for planar five-point one-mass processes},
  \href{http://dx.doi.org/10.1007/jhep11(2020)117}{\emph{Journal of High Energy
  Physics} {\bf 2020} (Nov, 2020) }.

\bibitem{Adams_2018}
L.~Adams and S.~Weinzierl, \emph{The $\varepsilon$-form of the differential
  equations for feynman integrals in the elliptic case},
  \href{http://dx.doi.org/10.1016/j.physletb.2018.04.002}{\emph{Physics Letters
  B} {\bf 781} (Jun, 2018) 270?278}.

\bibitem{BOGNER2017528}
C.~Bogner, A.~Schweitzer and S.~Weinzierl, \emph{Analytic continuation and
  numerical evaluation of the kite integral and the equal mass sunrise
  integral},
  \href{http://dx.doi.org/https://doi.org/10.1016/j.nuclphysb.2017.07.008}{\emph{Nuclear
  Physics B} {\bf 922} (2017) 528--550}.

\bibitem{Broedel:2019kmn}
J.~Broedel, C.~Duhr, F.~Dulat, R.~Marzucca, B.~Penante and L.~Tancredi,
  \emph{{An analytic solution for the equal-mass banana graph}},
  \href{http://dx.doi.org/10.1007/JHEP09(2019)112}{\emph{JHEP} {\bf 09} (2019)
  112}, [\href{http://arxiv.org/abs/1907.03787}{{\tt 1907.03787}}].

\bibitem{Abreu:2019fgk}
S.~Abreu, M.~Becchetti, C.~Duhr and R.~Marzucca, \emph{{Three-loop
  contributions to the $\rho$ parameter and iterated integrals of modular
  forms}}, \href{http://dx.doi.org/10.1007/JHEP02(2020)050}{\emph{JHEP} {\bf
  02} (2020) 050}, [\href{http://arxiv.org/abs/1912.02747}{{\tt 1912.02747}}].

\bibitem{Bonciani:2008az}
R.~Bonciani, A.~Ferroglia, T.~Gehrmann, D.~Maitre and C.~Studerus,
  \emph{{Two-Loop Fermionic Corrections to Heavy-Quark Pair Production: The
  Quark-Antiquark Channel}},
  \href{http://dx.doi.org/10.1088/1126-6708/2008/07/129}{\emph{JHEP} {\bf 07}
  (2008) 129}, [\href{http://arxiv.org/abs/0806.2301}{{\tt 0806.2301}}].

\bibitem{Bonciani:2009nb}
R.~Bonciani, A.~Ferroglia, T.~Gehrmann and C.~Studerus, \emph{{Two-Loop Planar
  Corrections to Heavy-Quark Pair Production in the Quark-Antiquark Channel}},
  \href{http://dx.doi.org/10.1088/1126-6708/2009/08/067}{\emph{JHEP} {\bf 08}
  (2009) 067}, [\href{http://arxiv.org/abs/0906.3671}{{\tt 0906.3671}}].

\bibitem{Bonciani:2010mn}
R.~Bonciani, A.~Ferroglia, T.~Gehrmann, A.~von Manteuffel and C.~Studerus,
  \emph{{Two-Loop Leading Color Corrections to Heavy-Quark Pair Production in
  the Gluon Fusion Channel}},
  \href{http://dx.doi.org/10.1007/JHEP01(2011)102}{\emph{JHEP} {\bf 01} (2011)
  102}, [\href{http://arxiv.org/abs/1011.6661}{{\tt 1011.6661}}].

\bibitem{Bonciani:2013ywa}
R.~Bonciani, A.~Ferroglia, T.~Gehrmann, A.~von Manteuffel and C.~Studerus,
  \emph{{Light-quark two-loop corrections to heavy-quark pair production in the
  gluon fusion channel}},
  \href{http://dx.doi.org/10.1007/JHEP12(2013)038}{\emph{JHEP} {\bf 12} (2013)
  038}, [\href{http://arxiv.org/abs/1309.4450}{{\tt 1309.4450}}].

\bibitem{vonManteuffel:2013uoa}
A.~von Manteuffel and C.~Studerus, \emph{{Massive planar and non-planar double
  box integrals for light Nf contributions to gg-\ensuremath{>}tt}},
  \href{http://dx.doi.org/10.1007/JHEP10(2013)037}{\emph{JHEP} {\bf 10} (2013)
  037}, [\href{http://arxiv.org/abs/1306.3504}{{\tt 1306.3504}}].

\bibitem{DiVita:2018nnh}
S.~Di~Vita, S.~Laporta, P.~Mastrolia, A.~Primo and U.~Schubert, \emph{{Master
  integrals for the NNLO virtual corrections to $\mu e$ scattering in QED: the
  non-planar graphs}},
  \href{http://dx.doi.org/10.1007/JHEP09(2018)016}{\emph{JHEP} {\bf 09} (2018)
  016}, [\href{http://arxiv.org/abs/1806.08241}{{\tt 1806.08241}}].

\bibitem{Becchetti:2019tjy}
M.~Becchetti, R.~Bonciani, V.~Casconi, A.~Ferroglia, S.~Lavacca and A.~von
  Manteuffel, \emph{{Master Integrals for the two-loop, non-planar QCD
  corrections to top-quark pair production in the quark-annihilation channel}},
  \href{http://dx.doi.org/10.1007/JHEP08(2019)071}{\emph{JHEP} {\bf 08} (2019)
  071}, [\href{http://arxiv.org/abs/1904.10834}{{\tt 1904.10834}}].

\bibitem{DiVita:2019lpl}
S.~Di~Vita, T.~Gehrmann, S.~Laporta, P.~Mastrolia, A.~Primo and U.~Schubert,
  \emph{{Master integrals for the NNLO virtual corrections to $
  q\overline{q}\to t\overline{t} $ scattering in QCD: the non-planar graphs}},
  \href{http://dx.doi.org/10.1007/JHEP06(2019)117}{\emph{JHEP} {\bf 06} (2019)
  117}, [\href{http://arxiv.org/abs/1904.10964}{{\tt 1904.10964}}].

\bibitem{Wang:1981:PAU:800206.806398}
P.~S. Wang, \emph{A p-adic algorithm for univariate partial fractions},  in
  \emph{Proceedings of the Fourth ACM Symposium on Symbolic and Algebraic
  Computation}, SYMSAC '81, (New York, NY, USA), pp.~212--217, ACM, 1981.
\newblock \href{http://dx.doi.org/10.1145/800206.806398}{DOI}.

\bibitem{Wang:1982:PRR:1089292.1089293}
P.~S. Wang, M.~J.~T. Guy and J.~H. Davenport, \emph{P-adic reconstruction of
  rational numbers},
  \href{http://dx.doi.org/10.1145/1089292.1089293}{\emph{SIGSAM Bull.} {\bf 16}
  (May, 1982) 2--3}.

\bibitem{Trager:2006:1145768}
\emph{ISSAC '06: Proceedings of the 2006 International Symposium on Symbolic
  and Algebraic Computation}, (New York, NY, USA), ACM, 2006.

\bibitem{vonManteuffel:2014ixa}
A.~von Manteuffel and R.~M. Schabinger, \emph{{A novel approach to integration
  by parts reduction}},
  \href{http://dx.doi.org/10.1016/j.physletb.2015.03.029}{\emph{Phys. Lett.}
  {\bf B744} (2015) 101--104}, [\href{http://arxiv.org/abs/1406.4513}{{\tt
  1406.4513}}].

\bibitem{Peraro:2016wsq}
T.~Peraro, \emph{{Scattering amplitudes over finite fields and multivariate
  functional reconstruction}},
  \href{http://dx.doi.org/10.1007/JHEP12(2016)030}{\emph{JHEP} {\bf 12} (2016)
  030}, [\href{http://arxiv.org/abs/1608.01902}{{\tt 1608.01902}}].

\bibitem{Peraro:2019svx}
T.~Peraro, \emph{{FiniteFlow: multivariate functional reconstruction using
  finite fields and dataflow graphs}},
  \href{http://arxiv.org/abs/1905.08019}{{\tt 1905.08019}}.

\bibitem{Ellis:2008ir}
R.~Ellis, W.~T. Giele, Z.~Kunszt and K.~Melnikov, \emph{{Masses, fermions and
  generalized $D$-dimensional unitarity}},
  \href{http://dx.doi.org/10.1016/j.nuclphysb.2009.07.023}{\emph{Nucl. Phys. B}
  {\bf 822} (2009) 270--282}, [\href{http://arxiv.org/abs/0806.3467}{{\tt
  0806.3467}}].

\bibitem{Melnikov:2009dn}
K.~Melnikov and M.~Schulze, \emph{{NLO QCD corrections to top quark pair
  production and decay at hadron colliders}},
  \href{http://dx.doi.org/10.1088/1126-6708/2009/08/049}{\emph{JHEP} {\bf 08}
  (2009) 049}, [\href{http://arxiv.org/abs/0907.3090}{{\tt 0907.3090}}].

\bibitem{Badger:2011yu}
S.~Badger, R.~Sattler and V.~Yundin, \emph{{One-Loop Helicity Amplitudes for
  $t\bar{t}$ Production at Hadron Colliders}},
  \href{http://dx.doi.org/10.1103/PhysRevD.83.074020}{\emph{Phys. Rev. D} {\bf
  83} (2011) 074020}, [\href{http://arxiv.org/abs/1101.5947}{{\tt 1101.5947}}].

\bibitem{Badger:2017gta}
S.~Badger, C.~Brønnum-Hansen, F.~Buciuni and D.~O'Connell, \emph{{A unitarity
  compatible approach to one-loop amplitudes with massive fermions}},
  \href{http://dx.doi.org/10.1007/JHEP06(2017)141}{\emph{JHEP} {\bf 06} (2017)
  141}, [\href{http://arxiv.org/abs/1703.05734}{{\tt 1703.05734}}].

\bibitem{Bernreuther:2001rq}
W.~Bernreuther, A.~Brandenburg, Z.~G. Si and P.~Uwer, \emph{{Top quark spin
  correlations at hadron colliders: Predictions at next-to-leading order QCD}},
  \href{http://dx.doi.org/10.1103/PhysRevLett.87.242002}{\emph{Phys. Rev.
  Lett.} {\bf 87} (2001) 242002},
  [\href{http://arxiv.org/abs/hep-ph/0107086}{{\tt hep-ph/0107086}}].

\bibitem{Bernreuther:2004jv}
W.~Bernreuther, A.~Brandenburg, Z.~G. Si and P.~Uwer, \emph{{Top quark pair
  production and decay at hadron colliders}},
  \href{http://dx.doi.org/10.1016/j.nuclphysb.2004.04.019}{\emph{Nucl. Phys. B}
  {\bf 690} (2004) 81--137}, [\href{http://arxiv.org/abs/hep-ph/0403035}{{\tt
  hep-ph/0403035}}].

\bibitem{Bern:1993kr}
Z.~Bern, L.~J. Dixon and D.~A. Kosower, \emph{{Dimensionally regulated pentagon
  integrals}},
  \href{http://dx.doi.org/10.1016/0550-3213(94)90398-0}{\emph{Nucl. Phys. B}
  {\bf 412} (1994) 751--816}, [\href{http://arxiv.org/abs/hep-ph/9306240}{{\tt
  hep-ph/9306240}}].

\bibitem{Kotikov:1990kg}
A.~V. Kotikov, \emph{{Differential equations method: New technique for massive
  Feynman diagrams calculation}},
  \href{http://dx.doi.org/10.1016/0370-2693(91)90413-K}{\emph{Phys. Lett. B}
  {\bf 254} (1991) 158--164}.

\bibitem{Remiddi:1997ny}
E.~Remiddi, \emph{{Differential equations for Feynman graph amplitudes}},
  {\emph{Nuovo Cim. A} {\bf 110} (1997) 1435--1452},
  [\href{http://arxiv.org/abs/hep-th/9711188}{{\tt hep-th/9711188}}].

\bibitem{Gehrmann:1999as}
T.~Gehrmann and E.~Remiddi, \emph{{Differential equations for two loop four
  point functions}},
  \href{http://dx.doi.org/10.1016/S0550-3213(00)00223-6}{\emph{Nucl. Phys. B}
  {\bf 580} (2000) 485--518}, [\href{http://arxiv.org/abs/hep-ph/9912329}{{\tt
  hep-ph/9912329}}].

\bibitem{Henn:2013pwa}
J.~M. Henn, \emph{{Multiloop integrals in dimensional regularization made
  simple}}, \href{http://dx.doi.org/10.1103/PhysRevLett.110.251601}{\emph{Phys.
  Rev. Lett.} {\bf 110} (2013) 251601},
  [\href{http://arxiv.org/abs/1304.1806}{{\tt 1304.1806}}].

\bibitem{Catani:1998bh}
S.~Catani, \emph{{The Singular behavior of QCD amplitudes at two loop order}},
  \href{http://dx.doi.org/10.1016/S0370-2693(98)00332-3}{\emph{Phys. Lett.}
  {\bf B427} (1998) 161--171}, [\href{http://arxiv.org/abs/hep-ph/9802439}{{\tt
  hep-ph/9802439}}].

\bibitem{Becher:2009cu}
T.~Becher and M.~Neubert, \emph{{Infrared singularities of scattering
  amplitudes in perturbative QCD}},
  \href{http://dx.doi.org/10.1103/PhysRevLett.102.162001,
  10.1103/PhysRevLett.111.199905}{\emph{Phys. Rev. Lett.} {\bf 102} (2009)
  162001}, [\href{http://arxiv.org/abs/0901.0722}{{\tt 0901.0722}}].

\bibitem{Becher:2009qa}
T.~Becher and M.~Neubert, \emph{{On the Structure of Infrared Singularities of
  Gauge-Theory Amplitudes}},
  \href{http://dx.doi.org/10.1088/1126-6708/2009/06/081,
  10.1007/JHEP11(2013)024}{\emph{JHEP} {\bf 06} (2009) 081},
  [\href{http://arxiv.org/abs/0903.1126}{{\tt 0903.1126}}].

\bibitem{Gardi:2009qi}
E.~Gardi and L.~Magnea, \emph{{Factorization constraints for soft anomalous
  dimensions in QCD scattering amplitudes}},
  \href{http://dx.doi.org/10.1088/1126-6708/2009/03/079}{\emph{JHEP} {\bf 03}
  (2009) 079}, [\href{http://arxiv.org/abs/0901.1091}{{\tt 0901.1091}}].

\bibitem{Gardi:2009zv}
E.~Gardi and L.~Magnea, \emph{{Infrared singularities in QCD amplitudes}},
  \href{http://dx.doi.org/10.1393/ncc/i2010-10528-x}{\emph{Frascati Phys. Ser.}
  {\bf 50} (2010) 137--157}, [\href{http://arxiv.org/abs/0908.3273}{{\tt
  0908.3273}}].

\bibitem{Catani:2000ef}
S.~Catani, S.~Dittmaier and Z.~Trocsanyi, \emph{{One loop singular behavior of
  QCD and SUSY QCD amplitudes with massive partons}},
  \href{http://dx.doi.org/10.1016/S0370-2693(01)00065-X}{\emph{Phys. Lett. B}
  {\bf 500} (2001) 149--160}, [\href{http://arxiv.org/abs/hep-ph/0011222}{{\tt
  hep-ph/0011222}}].

\bibitem{Ferroglia:2009ep}
A.~Ferroglia, M.~Neubert, B.~D. Pecjak and L.~L. Yang, \emph{{Two-loop
  divergences of scattering amplitudes with massive partons}},
  \href{http://dx.doi.org/10.1103/PhysRevLett.103.201601}{\emph{Phys. Rev.
  Lett.} {\bf 103} (2009) 201601}, [\href{http://arxiv.org/abs/0907.4791}{{\tt
  0907.4791}}].

\bibitem{Ferroglia:2009ii}
A.~Ferroglia, M.~Neubert, B.~D. Pecjak and L.~L. Yang, \emph{{Two-loop
  divergences of massive scattering amplitudes in non-abelian gauge theories}},
  \href{http://dx.doi.org/10.1088/1126-6708/2009/11/062}{\emph{JHEP} {\bf 11}
  (2009) 062}, [\href{http://arxiv.org/abs/0908.3676}{{\tt 0908.3676}}].

\bibitem{Kleiss:1985yh}
R.~Kleiss and W.~J. Stirling, \emph{{Spinor Techniques for Calculating p anti-p
  ---\ensuremath{>} W+- / Z0 + Jets}},
  \href{http://dx.doi.org/10.1016/0550-3213(85)90285-8}{\emph{Nucl. Phys. B}
  {\bf 262} (1985) 235--262}.

\bibitem{Hodges:2009hk}
A.~Hodges, \emph{{Eliminating spurious poles from gauge-theoretic amplitudes}},
  \href{http://dx.doi.org/10.1007/JHEP05(2013)135}{\emph{JHEP} {\bf 05} (2013)
  135}, [\href{http://arxiv.org/abs/0905.1473}{{\tt 0905.1473}}].

\bibitem{Rodrigo:2005eu}
G.~Rodrigo, \emph{{Multigluonic scattering amplitudes of heavy quarks}},
  \href{http://dx.doi.org/10.1088/1126-6708/2005/09/079}{\emph{JHEP} {\bf 09}
  (2005) 079}, [\href{http://arxiv.org/abs/hep-ph/0508138}{{\tt
  hep-ph/0508138}}].

\bibitem{Schwinn:2007ee}
C.~Schwinn and S.~Weinzierl, \emph{{On-shell recursion relations for all Born
  QCD amplitudes}},
  \href{http://dx.doi.org/10.1088/1126-6708/2007/04/072}{\emph{JHEP} {\bf 04}
  (2007) 072}, [\href{http://arxiv.org/abs/hep-ph/0703021}{{\tt
  hep-ph/0703021}}].

\bibitem{Arkani-Hamed:2017jhn}
N.~Arkani-Hamed, T.-C. Huang and Y.-t. Huang, \emph{{Scattering Amplitudes For
  All Masses and Spins}},  \href{http://arxiv.org/abs/1709.04891}{{\tt
  1709.04891}}.

\bibitem{Badger:2013gxa}
S.~Badger, H.~Frellesvig and Y.~Zhang, \emph{{A Two-Loop Five-Gluon Helicity
  Amplitude in QCD}},
  \href{http://dx.doi.org/10.1007/JHEP12(2013)045}{\emph{JHEP} {\bf 12} (2013)
  045}, [\href{http://arxiv.org/abs/1310.1051}{{\tt 1310.1051}}].

\bibitem{Abreu:2020xvt}
S.~Abreu, J.~Dormans, F.~Febres~Cordero, H.~Ita, M.~Kraus, B.~Page et~al.,
  \emph{{Caravel: A C++ Framework for the Computation of Multi-Loop Amplitudes
  with Numerical Unitarity}},  \href{http://arxiv.org/abs/2009.11957}{{\tt
  2009.11957}}.

\bibitem{Chen:2019wyb}
L.~Chen, \emph{{A prescription for projectors to compute helicity amplitudes in
  D dimensions}},  \href{http://arxiv.org/abs/1904.00705}{{\tt 1904.00705}}.

\bibitem{Peraro:2019cjj}
T.~Peraro and L.~Tancredi, \emph{{Physical projectors for multi-leg helicity
  amplitudes}}, \href{http://dx.doi.org/10.1007/JHEP07(2019)114}{\emph{JHEP}
  {\bf 07} (2019) 114}, [\href{http://arxiv.org/abs/1906.03298}{{\tt
  1906.03298}}].

\bibitem{Peraro:2020sfm}
T.~Peraro and L.~Tancredi, \emph{{Tensor decomposition for bosonic and
  fermionic scattering amplitudes}},
  \href{http://arxiv.org/abs/2012.00820}{{\tt 2012.00820}}.

\bibitem{Nogueira:1991ex}
P.~Nogueira, \emph{{Automatic Feynman graph generation}},
  \href{http://dx.doi.org/10.1006/jcph.1993.1074}{\emph{J. Comput. Phys.} {\bf
  105} (1993) 279--289}.

\bibitem{Kuipers:2012rf}
J.~Kuipers, T.~Ueda, J.~A.~M. Vermaseren and J.~Vollinga, \emph{{FORM version
  4.0}}, \href{http://dx.doi.org/10.1016/j.cpc.2012.12.028}{\emph{Comput. Phys.
  Commun.} {\bf 184} (2013) 1453--1467},
  [\href{http://arxiv.org/abs/1203.6543}{{\tt 1203.6543}}].

\bibitem{Ruijl:2017dtg}
B.~Ruijl, T.~Ueda and J.~Vermaseren, \emph{{FORM version 4.2}},
  \href{http://arxiv.org/abs/1707.06453}{{\tt 1707.06453}}.

\bibitem{Cullen:2010jv}
G.~Cullen, M.~Koch-Janusz and T.~Reiter, \emph{{Spinney: A Form Library for
  Helicity Spinors}},
  \href{http://dx.doi.org/10.1016/j.cpc.2011.06.007}{\emph{Comput. Phys.
  Commun.} {\bf 182} (2011) 2368--2387},
  [\href{http://arxiv.org/abs/1008.0803}{{\tt 1008.0803}}].

\bibitem{Hartanto:2019uvl}
H.~B. Hartanto, S.~Badger, C.~Brønnum-Hansen and T.~Peraro, \emph{{A numerical
  evaluation of planar two-loop helicity amplitudes for a W-boson plus four
  partons}}, \href{http://dx.doi.org/10.1007/JHEP09(2019)119}{\emph{JHEP} {\bf
  09} (2019) 119}, [\href{http://arxiv.org/abs/1906.11862}{{\tt 1906.11862}}].

\bibitem{Mastrolia:2016dhn}
P.~Mastrolia, T.~Peraro and A.~Primo, \emph{{Adaptive Integrand Decomposition
  in parallel and orthogonal space}},
  \href{http://dx.doi.org/10.1007/JHEP08(2016)164}{\emph{JHEP} {\bf 08} (2016)
  164}, [\href{http://arxiv.org/abs/1605.03157}{{\tt 1605.03157}}].

\bibitem{Badger:2017jhb}
S.~Badger, C.~Br\o{}nnum-Hansen, H.~B. Hartanto and T.~Peraro, \emph{{First
  look at two-loop five-gluon scattering in QCD}},
  \href{http://dx.doi.org/10.1103/PhysRevLett.120.092001}{\emph{Phys. Rev.
  Lett.} {\bf 120} (2018) 092001}, [\href{http://arxiv.org/abs/1712.02229}{{\tt
  1712.02229}}].

\bibitem{Badger:2018enw}
S.~Badger, C.~Br\o{}nnum-Hansen, H.~B. Hartanto and T.~Peraro, \emph{{Analytic
  helicity amplitudes for two-loop five-gluon scattering: the single-minus
  case}}, \href{http://dx.doi.org/10.1007/JHEP01(2019)186}{\emph{JHEP} {\bf 01}
  (2019) 186}, [\href{http://arxiv.org/abs/1811.11699}{{\tt 1811.11699}}].

\bibitem{Badger:2021nhg}
S.~Badger, H.~B. Hartanto and S.~Zoia, \emph{{Two-loop QCD corrections to
  $Wb\bar{b}$ production at hadron colliders}},
  \href{http://arxiv.org/abs/2102.02516}{{\tt 2102.02516}}.

\bibitem{Lee:2012cn}
R.~N. Lee, \emph{{Presenting LiteRed: a tool for the Loop InTEgrals
  REDuction}},  \href{http://arxiv.org/abs/1212.2685}{{\tt 1212.2685}}.

\bibitem{Laporta:2001dd}
S.~Laporta, \emph{{High precision calculation of multiloop Feynman integrals by
  difference equations}},
  \href{http://dx.doi.org/10.1016/S0217-751X(00)00215-7,
  10.1142/S0217751X00002157}{\emph{Int. J. Mod. Phys.} {\bf A15} (2000)
  5087--5159}, [\href{http://arxiv.org/abs/hep-ph/0102033}{{\tt
  hep-ph/0102033}}].

\bibitem{goncharov2011multiple}
A.~B. Goncharov, \emph{Multiple polylogarithms, cyclotomy and modular
  complexes},  2011.

\bibitem{Moch_2002}
S.~Moch, P.~Uwer and S.~Weinzierl, \emph{Nested sums, expansion of
  transcendental functions, and multiscale multiloop integrals},
  \href{http://dx.doi.org/10.1063/1.1471366}{\emph{Journal of Mathematical
  Physics} {\bf 43} (Jun, 2002) 3363?3386}.

\bibitem{Borwein:1999js}
J.~M. Borwein, D.~M. Bradley, D.~J. Broadhurst and P.~Lisonek, \emph{{Special
  values of multiple polylogarithms}},
  \href{http://dx.doi.org/10.1090/S0002-9947-00-02616-7}{\emph{Trans. Am. Math.
  Soc.} {\bf 353} (2001) 907--941},
  [\href{http://arxiv.org/abs/math/9910045}{{\tt math/9910045}}].

\bibitem{Mastrolia:2017pfy}
P.~Mastrolia, M.~Passera, A.~Primo and U.~Schubert, \emph{{Master integrals for
  the NNLO virtual corrections to $\mu e$ scattering in QED: the planar
  graphs}}, \href{http://dx.doi.org/10.1007/JHEP11(2017)198}{\emph{JHEP} {\bf
  11} (2017) 198}, [\href{http://arxiv.org/abs/1709.07435}{{\tt 1709.07435}}].

\bibitem{Chen:2019zoy}
L.-B. Chen and J.~Wang, \emph{{Master integrals of a planar double-box family
  for top-quark pair production}},
  \href{http://dx.doi.org/10.1016/j.physletb.2019.03.030}{\emph{Phys. Lett. B}
  {\bf 792} (2019) 50--55}, [\href{http://arxiv.org/abs/1903.04320}{{\tt
  1903.04320}}].

\bibitem{Chen:1977oja}
K.-T. Chen, \emph{{Iterated path integrals}},
  \href{http://dx.doi.org/10.1090/S0002-9904-1977-14320-6}{\emph{Bull. Am.
  Math. Soc.} {\bf 83} (1977) 831--879}.

\bibitem{Brown:2011}
F.~Brown, \emph{Iterated integrals in quantum field theory},
  \href{http://dx.doi.org/10.1017/CBO9781139208642.006}{\emph{Geometric and
  Topological Methods for Quantum Field Theory} (01, 2011) }.

\bibitem{Duhr:2019tlz}
C.~Duhr and F.~Dulat, \emph{{PolyLogTools \textemdash{} polylogs for the
  masses}}, \href{http://dx.doi.org/10.1007/JHEP08(2019)135}{\emph{JHEP} {\bf
  08} (2019) 135}, [\href{http://arxiv.org/abs/1904.07279}{{\tt 1904.07279}}].

\bibitem{Panzer:2015ida}
E.~Panzer, \emph{{Feynman integrals and hyperlogarithms}}.
\newblock PhD thesis, Humboldt U., 2015.
\newblock \href{http://arxiv.org/abs/1506.07243}{{\tt 1506.07243}}.
\newblock 10.18452/17157.

\bibitem{Bauer:2000cp}
C.~W. Bauer, A.~Frink and R.~Kreckel, \emph{{Introduction to the GiNaC
  framework for symbolic computation within the C++ programming language}},
  \href{http://dx.doi.org/10.1006/jsco.2001.0494}{\emph{J. Symb. Comput.} {\bf
  33} (2002) 1--12}, [\href{http://arxiv.org/abs/cs/0004015}{{\tt
  cs/0004015}}].

\bibitem{silverman}
J.~H. Silverman, \emph{{The Arithmetic of Elliptic Curves}}.
\newblock Springer, 2nd~ed., 1986.

\bibitem{Broedel:2017kkb}
J.~Broedel, C.~Duhr, F.~Dulat and L.~Tancredi, \emph{{Elliptic polylogarithms
  and iterated integrals on elliptic curves. Part I: general formalism}},
  \href{http://dx.doi.org/10.1007/JHEP05(2018)093}{\emph{JHEP} {\bf 05} (2018)
  093}, [\href{http://arxiv.org/abs/1712.07089}{{\tt 1712.07089}}].

\bibitem{Frellesvig_2017}
H.~Frellesvig and C.~G. Papadopoulos, \emph{Cuts of feynman integrals in baikov
  representation},
  \href{http://dx.doi.org/10.1007/jhep04(2017)083}{\emph{Journal of High Energy
  Physics} {\bf 2017} (Apr, 2017) }.

\bibitem{Primo:2016ebd}
A.~Primo and L.~Tancredi, \emph{{On the maximal cut of Feynman integrals and
  the solution of their differential equations}},
  \href{http://dx.doi.org/10.1016/j.nuclphysb.2016.12.021}{\emph{Nucl. Phys. B}
  {\bf 916} (2017) 94--116}, [\href{http://arxiv.org/abs/1610.08397}{{\tt
  1610.08397}}].

\bibitem{Primo:2017ipr}
A.~Primo and L.~Tancredi, \emph{{Maximal cuts and differential equations for
  Feynman integrals. An application to the three-loop massive banana graph}},
  \href{http://dx.doi.org/10.1016/j.nuclphysb.2017.05.018}{\emph{Nucl. Phys. B}
  {\bf 921} (2017) 316--356}, [\href{http://arxiv.org/abs/1704.05465}{{\tt
  1704.05465}}].

\bibitem{brown2017multiple}
F.~Brown, \emph{Multiple modular values and the relative completion of the
  fundamental group of $m_{1,1}$},  2017.

\bibitem{Broedel:2017siw}
J.~Broedel, C.~Duhr, F.~Dulat and L.~Tancredi, \emph{{Elliptic polylogarithms
  and iterated integrals on elliptic curves II: an application to the sunrise
  integral}}, \href{http://dx.doi.org/10.1103/PhysRevD.97.116009}{\emph{Phys.
  Rev. D} {\bf 97} (2018) 116009}, [\href{http://arxiv.org/abs/1712.07095}{{\tt
  1712.07095}}].

\bibitem{Broedel:2018iwv}
J.~Broedel, C.~Duhr, F.~Dulat, B.~Penante and L.~Tancredi, \emph{{Elliptic
  symbol calculus: from elliptic polylogarithms to iterated integrals of
  Eisenstein series}},
  \href{http://dx.doi.org/10.1007/JHEP08(2018)014}{\emph{JHEP} {\bf 08} (2018)
  014}, [\href{http://arxiv.org/abs/1803.10256}{{\tt 1803.10256}}].

\bibitem{Duhr:2019rrs}
C.~Duhr and L.~Tancredi, \emph{{Algorithms and tools for iterated Eisenstein
  integrals}}, \href{http://dx.doi.org/10.1007/JHEP02(2020)105}{\emph{JHEP}
  {\bf 02} (2020) 105}, [\href{http://arxiv.org/abs/1912.00077}{{\tt
  1912.00077}}].

\bibitem{Vollinga_2005}
J.~Vollinga and S.~Weinzierl, \emph{Numerical evaluation of multiple
  polylogarithms},
  \href{http://dx.doi.org/10.1016/j.cpc.2004.12.009}{\emph{Computer Physics
  Communications} {\bf 167} (May, 2005) 177?194}.

\bibitem{Smirnov:2015mct}
A.~V. Smirnov, \emph{{FIESTA4: Optimized Feynman integral calculations with GPU
  support}}, \href{http://dx.doi.org/10.1016/j.cpc.2016.03.013}{\emph{Comput.
  Phys. Commun.} {\bf 204} (2016) 189--199},
  [\href{http://arxiv.org/abs/1511.03614}{{\tt 1511.03614}}].

\bibitem{Borowka:2017idc}
S.~Borowka, G.~Heinrich, S.~Jahn, S.~P. Jones, M.~Kerner, J.~Schlenk et~al.,
  \emph{{pySecDec: a toolbox for the numerical evaluation of multi-scale
  integrals}},  \href{http://arxiv.org/abs/1703.09692}{{\tt 1703.09692}}.

\bibitem{Melnikov:2000zc}
K.~Melnikov and T.~van Ritbergen, \emph{{The Three loop on-shell
  renormalization of QCD and QED}},
  \href{http://dx.doi.org/10.1016/S0550-3213(00)00526-5}{\emph{Nucl. Phys. B}
  {\bf 591} (2000) 515--546}, [\href{http://arxiv.org/abs/hep-ph/0005131}{{\tt
  hep-ph/0005131}}].

\bibitem{Czakon:2007ej}
M.~Czakon, A.~Mitov and S.~Moch, \emph{{Heavy-quark production in massless
  quark scattering at two loops in QCD}},
  \href{http://dx.doi.org/10.1016/j.physletb.2007.06.020}{\emph{Phys. Lett. B}
  {\bf 651} (2007) 147--159}, [\href{http://arxiv.org/abs/0705.1975}{{\tt
  0705.1975}}].

\bibitem{Czakon:2007wk}
M.~Czakon, A.~Mitov and S.~Moch, \emph{{Heavy-quark production in gluon fusion
  at two loops in QCD}},
  \href{http://dx.doi.org/10.1016/j.nuclphysb.2008.02.001}{\emph{Nucl. Phys. B}
  {\bf 798} (2008) 210--250}, [\href{http://arxiv.org/abs/0707.4139}{{\tt
  0707.4139}}].

\end{thebibliography}\endgroup

\end{document}